\documentclass{aa}  
\usepackage{graphicx}
\usepackage{txfonts}
\usepackage[colorlinks=true,citecolor=blue,linkcolor=blue,urlcolor=cyan]{hyperref}
\usepackage{pdflscape}
\newcommand{\nodata}{}
\renewcommand\labelitemii{$\m@th\bullet$}
\begin{document} 

\title{Lupus DANCe}
\subtitle{Census of stars and 6D structure with Gaia-DR2 data\thanks{Tables \ref{tab_members}, \ref{tab_prob}, \ref{tab_isochrone}, and \ref{tab_teff_age} are only available in electronic form at the CDS via anonymous ftp to cdsarc.u-strasbg.fr (130.79.128.5) or via http://cdsweb.u-strasbg.fr/cgi-bin/qcat?J/A+A/}}

\author{
P.A.B.~Galli \inst{1}
\and
H.~Bouy \inst{1}
\and
J.~Olivares\inst{1}
\and 
N.~Miret-Roig\inst{1}
\and
R.G.~Vieira\inst{2}
\and
L.M.~Sarro\inst{3}
\and
D.~Barrado\inst{4}
\and
A.~Berihuete\inst{5}
\and
C.~Bertout \inst{6}
\and 
E.~Bertin \inst{6}
\and J.-C.~Cuillandre \inst{7}
}

\institute{
Laboratoire d’Astrophysique de Bordeaux, Univ. Bordeaux, CNRS, B18N, allée Geoffroy Saint-Hillaire, F-33615 Pessac, France\\
\email{phillip.galli@u-bordeaux.fr}
\and
Departamento de F\'isica, Universidade Federal de Sergipe, S\~ao Crist\'ov\~ao, SE, Brazil
\and
Depto. de Inteligencia Artificial, UNED, Juan del Rosal, 16, 28040 Madrid, Spain
\and
Centro de Astrobiolog\'ia, Depto. de Astrof\'isica, INTA-CSIC, ESAC Campus, Camino Bajo del Castillo s/n, 28692 Villanueva de la Ca\~nada, Madrid, Spain
\and
Dept. Statistics and Operations Research, University of C\'adiz, Campus Universitario R\'io San Pedro s/n, 11510 Puerto Real, \\C\'adiz, Spain
\and
Institut d’Astrophysique de Paris, CNRS UMR 7095 and UPMC, 98bis bd Arago, 75014 Paris, France
\and
AIM Paris Saclay, CNRS/INSU, CEA/Irfu, Université Paris Diderot, Orme des Merisiers, France
}

\date{Received XXX; accepted XXX}

 \abstract
{Lupus is recognised as one of the closest star-forming regions, but the lack of trigonometric parallaxes in the pre-\textit{Gaia} era hampered many studies on the kinematic properties of this region and led to incomplete censuses of its stellar population.    }
{We use the second data release of the \textit{Gaia} space mission combined with published ancillary radial velocity data to revise the census of stars and investigate the 6D structure of the Lupus complex. }
{We performed a new membership analysis of the Lupus association based on astrometric and photometric data over a field of 160~deg$^{2}$ around the main molecular clouds of the complex and compared the properties of the various subgroups in this region. }
{We identified 137 high-probability members of the Lupus association of young stars, including 47 stars that had never been reported as members before. Many of the historically known stars associated with the Lupus region identified in previous studies are more likely to be field stars or members of the adjacent Scorpius-Centaurus association. Our new sample of members covers the magnitude and mass range from $G\simeq8$ to $G\simeq18$~mag and from 0.03 to 2.4~M$_{\odot}$, respectively. We compared the kinematic properties of the stars projected towards the molecular clouds Lupus~1 to 6 and showed that these subgroups are located at roughly the same distance (about 160~pc) and move with the same spatial velocity. Our age estimates inferred from stellar models show that the Lupus subgroups are coeval (with median ages ranging from about 1 to 3~Myr). The Lupus association appears to be younger than the population of young stars in the Corona-Australis star-forming region recently investigated by our team using a similar methodology. The initial mass function of the Lupus association inferred from the distribution of spectral types shows little variation compared to other star-forming regions. 
}
{In this paper, we provide an updated sample of cluster members based on \textit{Gaia} data and construct the most complete picture of the 3D structure and 3D space motion of the Lupus complex.}

\keywords{open clusters and associations: individual: Lupus - Stars: formation - Stars: distances - Methods: statistical - Parallaxes - Proper motions}
\maketitle

\section{Introduction}\label{section1}

The Lupus molecular cloud complex is one of the closest and largest low-mass star-forming regions of the southern sky. It consists of several subgroups associated with different molecular clouds labelled as Lupus~1 to 9 that exhibit distinct morphologies as revealed by CO surveys and extinction maps \citep{Hara1999,Cambresy1999,Tachihara2001,Dobashi2005}. Although they are part of the same complex, Lupus clouds differ significantly regarding their star formation activity. Some clouds have dense concentrations of T~Tauri stars (e.g. Lupus 3) while others (e.g. Lupus~7, 8 and 9) display no signs of ongoing star formation \citep{Comeron2008}.  

Early surveys leading to the identification of Lupus stars mostly focused on the molecular clouds with active star formation (i.e. Lupus~1 to 4), but more recently a few young stellar objects (YSOs) have also been discovered in Lupus~5 and 6 \citep{Manara2018,Melton2020}. Most of the hitherto known classical T~Tauri stars (CTTSs) in Lupus were identified by \citet{Schwartz1977} and \citet{Hughes1994} based on their strong H$\alpha$ and infrared excess emissions. Succeeding studies identified a more dispersed and older population of weak-emission line T~Tauri stars (WTTSs) from ROSAT X-ray pointed observations surrounding the Lupus clouds, which greatly exceed the number of CTTSs located in the immediate vicinity of the molecular clouds \citep{Krautter1997,Wichmann1997b,Wichmann1997}. The census of Lupus stars was later expanded by \citet{Merin2008} based on infrared observations collected with the \textit{Spitzer Space Telescope} as part of the \textit{cores to disk (c2d)} legacy project. In a subsequent study, \citet{Comeron2009} revealed a new population of cool stars and brown dwarfs potentially associated with Lupus based on their spectral energy distributions (SEDs) derived from optical and infrared photometry. \citet{Lopez-Marti2011} confirmed most of these sources to be genuine members of the Lupus region based on their proper motions and showed the importance of using kinematic information (e.g. proper motions, parallaxes, and radial velocities) to assess membership. 

The distance to Lupus has undergone several revisions and has become a matter of debate in the recent decades although it has always been recognised as one of the closest star-forming regions to the Sun. Distance determinations in the literature range from 100~pc \citep{Knude1998} to 360~pc \citep{Knude2001}. \citet{Franco1990} and \citet{Hughes1993} estimated the distances of 165$\pm$15~pc and 140$\pm$20~pc, respectively, based on the interstellar reddening of field stars. \citet{Lombardi2008} used a more robust method based on 2MASS wide field extinction maps and derived the distance of 155$\pm$8~pc with a depth of $51^{+61}_{-35}$~pc, which the authors explained as the Lupus clouds being at different distances. It is important to mention that these distances are only average estimates based on indirect methods because distances for individual stars derived from trigonometric parallaxes existed (until recently) for only a few stars in this region. For example, \citet{Bertout1999} used trigonometric parallaxes of five stars from the \textit{Hipparcos} catalogue \citep{Hipparcos} to estimate the distance to Lupus. Three of them, which are associated with Lupus~1, 2 and 4 have a mean parallax of $\varpi=6.79\pm1.50$~mas that roughly defines a distance of $147^{+42}_{-27}$~pc. The other two stars, located in Lupus~3, have a mean parallax of $\varpi=4.38\pm0.67$~mas, which puts this cloud much farther away at a distance of $228^{+41}_{-30}$~pc. Alternatively, kinematic distances to individual stars in this region were derived in the past from the convergent point method under the assumption that the Lupus stars are comoving \citep{Makarov2007,Galli2013}. The resulting distances confirmed the important depth effects (i.e. distance variations along the line of sight) reported in previous studies. 

Lupus is located between the Upper Scorpius (US) and Upper Centaurus-Lupus (UCL) subgroups of the Scorpius-Centaurus (Sco-Cen) association, but despite the close proximity in the sky it represents a more recent star formation episode in this region \citep{Preibisch2008}. In this context, the existence of a more dispersed and older population of WTTSs near the Lupus clouds as reported in previous studies is not surprising, but their association with the younger population concentrated in the immediate vicinity of the molecular clouds needs to be confirmed. The second data release of the \textit{Gaia} space mission \citep[Gaia-DR2,][]{GaiaDR2} provides the best astrometry available to date to further assess the membership of the historically known members. When combined with radial velocity information these data can be used to constrain not only the 3D positions of the stars, but also their 3D space motions. The stellar content of the Sco-Cen association was recently revised by \citet{Damiani2019} based on Gaia-DR2 data, but a dedicated study of the Lupus clouds is still missing in the literature. The proper motions and parallaxes in the Gaia-DR2 catalogue are more precise by one or two orders of magnitude compared to the ground-based astrometry used in previous studies of the Lupus region \citep[see e.g.][]{Makarov2007,Lopez-Marti2011,Galli2013}, and they will therefore allow us to obtain a clean sample of members and a more accurate picture of the overall 6D structure of the complex. 

This paper is one in a series as part of the Dynamical Analysis of Nearby Clusters project \citep[DANCe,][]{Bouy2013}. Here, we investigate the census of stars and kinematic properties of the Lupus association of young stars based on Gaia-DR2 data. It is organised as follows. In Section~\ref{section2} we compile the list of Lupus stars from previous studies in the literature that is used in Section~\ref{section3} to perform a new membership analysis based on a novel methodology developed by our team \citep{Sarro2014,Olivares2019}. In Section~\ref{section4} we revisit several properties of the Lupus subgroups (e.g. distance, spatial velocity, age, spatial distribution, and initial mass function) based on our new sample of cluster members. Finally, we summarise our conclusions in Section~\ref{section5}.

\section{Sample of Lupus stars from previous studies}\label{section2}

\begin{figure*}
\begin{center}
\includegraphics[width=0.82\textwidth]{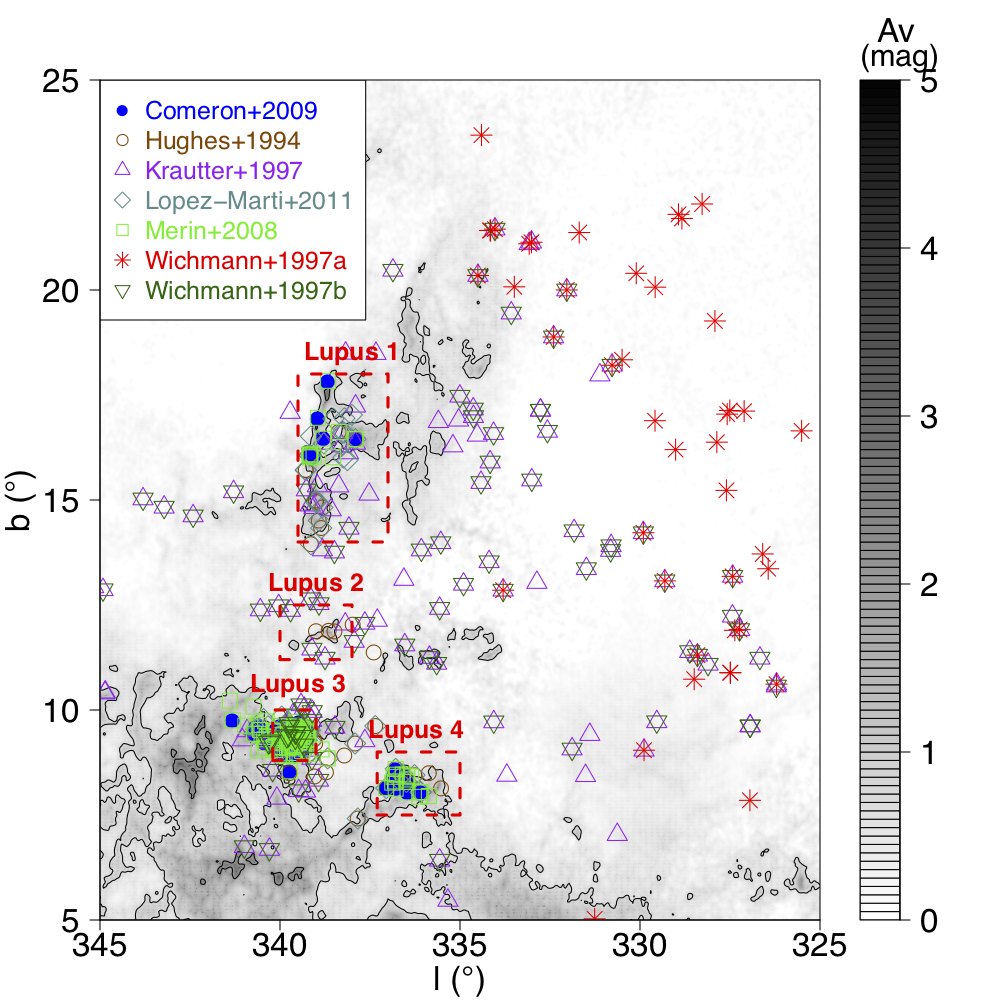}
\caption{Location of the Lupus stars identified in previous studies overlaid on the extinction map of \citet{Dobashi2005} in Galactic coordinates. Colours and symbols denote the samples of stars from each study. The red dashed lines indicate the position of the main molecular clouds defined by \citet{Hara1999}.
\label{fig_location_Lupus_lit} 
}
\end{center}
\end{figure*}

We construct our initial sample of stars based on the lists of members and candidate members that have been associated with the Lupus star-forming region in previous studies. We take the samples of (i) 73~stars from \citet{Hughes1994}, (ii) 136~stars from \citet{Krautter1997}, (iii) 92~stars from \citet{Wichmann1997b}, (iv) 48~stars from \citet{Wichmann1997}, (v) 159~stars from \citet{Merin2008}, (vi) 248 stars from Tables~4 and 9 of \citet{Comeron2009}, (vii) 82~stars from Tables~2 and 4 of \citet{Lopez-Marti2011}, and (viii) 69~stars from \citet{Damiani2019}. This results in a sample of 508~stars after removing the sources in common among the several lists of Lupus stars. We find proper motions and parallaxes for 441~stars of this sample in the Gaia-DR2 catalogue. Figure~\ref{fig_location_Lupus_lit} shows the location of the stars in this sample with respect to the main star-forming clouds of the Lupus complex. We use the boundaries defined in Figure~3 of \cite{Hara1999} to assign the sources in the Lupus sample to the corresponding molecular cloud of the complex based on their sky positions. This procedure does not confirm membership to the corresponding clouds, but allows us to compare the stellar proper motions of the stars projected against the various clouds. As discussed in Sect.~\ref{section1}, it is apparent that the stars identified by \citet{Krautter1997} and \citet{Wichmann1997b,Wichmann1997} constitute a more dispersed `off-cloud' population that surrounds the molecular clouds.

Similarly, we compile a list of known members of the adjacent Sco-Cen association that will be used in the forthcoming discussion. We restrict this sample to the to stars in the UCL subgroup that are located in the general region of the main star-forming clouds of the Lupus complex ($334^{\circ}<l<342^{\circ}$ and $5^{\circ}<b<25^{\circ}$). Our UCL sample is based on the lists of sources previously identified in this region by \citet{deZeeuw1999}, \citet{Preibisch2008}, \citet{Pecaut2016} and \citet{Damiani2019}. It includes1352~stars with measured proper motions and parallaxes in the Gaia-DR2 catalogue. 

\begin{figure}
\begin{center}
\sidecaption
\includegraphics[width=0.49\textwidth]{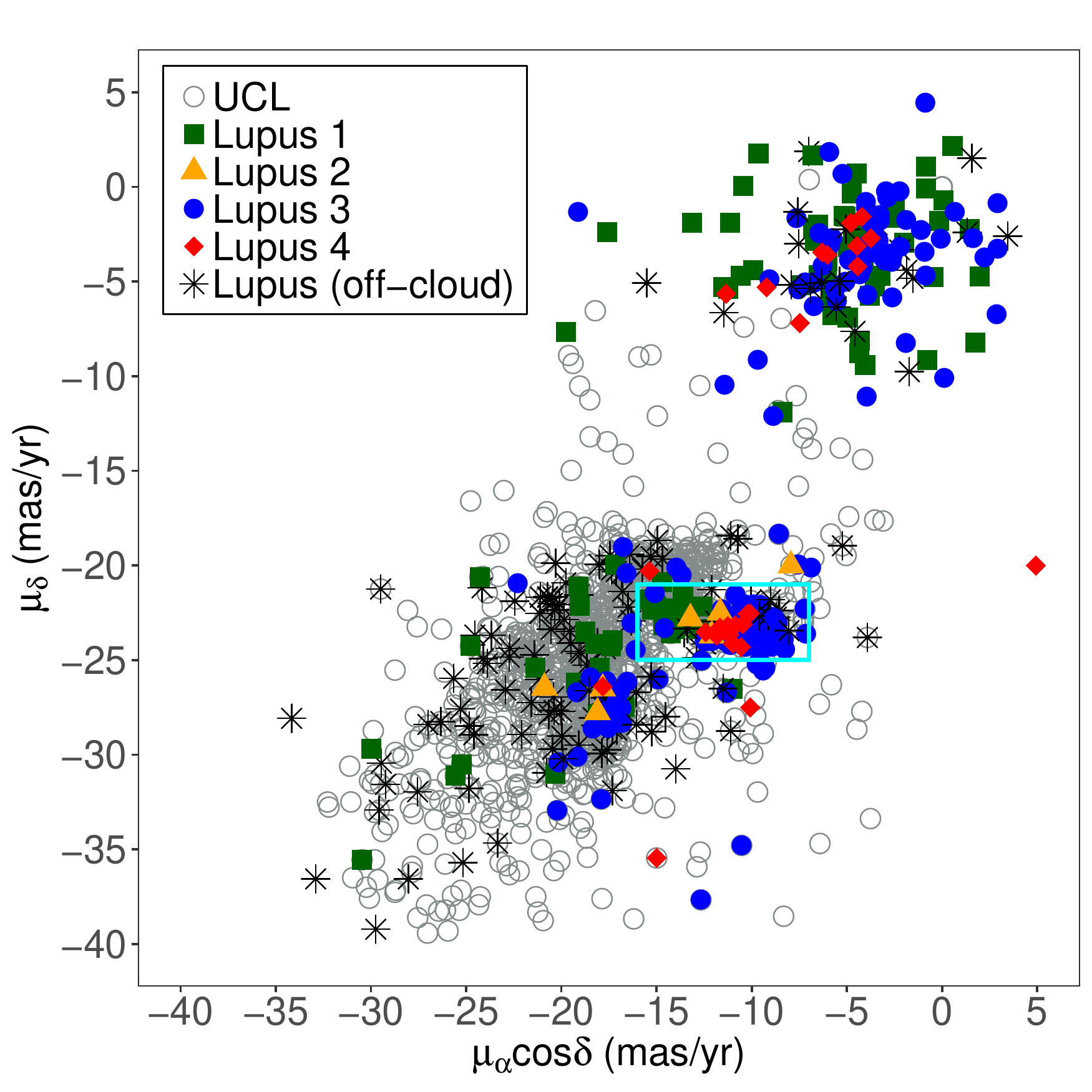}
\caption{Proper motion of Lupus and UCL stars identified in previous studies. The different colours denote the subgroups of the Lupus sample. The cyan rectangle indicates the locus of Lupus candidate stars in the space of proper motions (see discussion in Sect.~\ref{section2}).   
\label{fig_pm_Lupus_lit} 
}
\end{center}
\end{figure}

Figure~\ref{fig_pm_Lupus_lit} shows the distribution of proper motions for the Lupus and UCL stars compiled from the literature. This preliminary analysis reveals the existence of (at least) three populations of stars in the Lupus sample from the literature. First, there is a background population of sources projected towards Lupus~1, 3, and 4 with proper motions typically smaller than 10~mas/yr which are more likely to be unrelated to the Lupus star-forming region as anticipated in past studies \citep{Lopez-Marti2011,Galli2013}. Second, there is a more diffuse population of stars with proper motions that overlap with the UCL stars in this region of the sky. Indeed, many of them have been identified as Sco-Cen members in the recent study conducted by \citet{Damiani2019}. Finally, we note an overdensity of stars around $(\mu_{\alpha}\cos\delta,\mu_{\delta})\simeq(-11,-23)$~mas/yr (see highlighted region in Figure~\ref{fig_pm_Lupus_lit}) including stars projected towards the four molecular clouds of the complex (Lupus~1 to 4) and with proper motions consistent with membership in Lupus \citep[see e.g.][]{Galli2013}. The latter will be used in our forthcoming analysis (see Sect.~\ref{section3}) to search for additional members to the Lupus association and investigate the kinematic properties of this star-forming region. 

\section{Membership analysis}\label{section3}

The membership analysis performed in this study for the Lupus star-forming region is based on the methodology developed by \citet{Sarro2014}  and \citet{Olivares2019}. The main steps of our analysis are briefly summarised in this section and we refer the reader to the original papers for further details.   

First, we define the representation space of our membership analysis (i.e. the set of observables) that we use to classify the sources as cluster members or field stars. The representation space includes both astrometric and photometric features of the stars given in the Gaia-DR2 catalogue. It does not include sky positions since the Lupus association is spread over a relatively large sky region without any clear over-density with respect to the field population in terms of stellar positions (see Figure~\ref{fig_location_Lupus_lit}). In our case the inclusion of stellar positions in the representation dilutes the discriminant power of parallaxes, proper motions and photometry in the membership analysis, thus resulting in higher contamination rates. Furthermore, we do not use the blue photometry (BP) from Gaia-DR2 due to some inconsistencies reported in the literature in the BP system \citep[see e.g.][]{Apellaniz2018}. We measured the relative importance of the photometric features using a random-forest classifier and concluded that $G_{RP}$ is a more discriminant magnitude than $G$ for identifying members. So, our representation space for the membership analysis is defined by $\mu_{\alpha}\cos\delta$, $\mu_{\delta}$, $\varpi$, $G_{RP}$, and $G-G_{RP}$. 

The region of the sky downloaded from the Gaia-DR2 catalogue to perform our membership analysis is defined by $334^{\circ}\leq l\leq 342^{\circ}$ and $5^{\circ}\leq b\leq 25^{\circ}$ to include the molecular clouds of the complex with active star formation (Lupus~1 to 4). This input catalogue contains 14\,471\,847~sources (10\,470\,632 of them with complete data in the chosen representation space) whose membership status with respect to the Lupus region will be investigated in our analysis.

We modelled the field population using Gaussian Mixture Models (GMM)\footnote{A Gaussian Mixture Model is a model that describes a probability distribution as a linear combination of $k$ Gaussian distributions, where $k$ is the number of components.}. The field model was computed only once at the beginning of our membership analysis and held fixed during the process. We tested field models with 20, 40, 60, 80, 100, 120, 140 and 160 components, and inferred their parameters using a random sample of 10$^6$ sources from our input catalogue. We compute the Bayesian information criteria (BIC) for each one of the models and take the one (with 80~components) that returns the minimum BIC value as the optimum model for our analysis. The main objective our membership analysis is to distinguish the Lupus association from the rest of the sources in the region surveyed by our study which we collectively denote as field population. It includes background sources, foreground stars and stellar groups at similar distances of Lupus, but with different properties (e.g. the Sco-Cen subgroups). The number of GMM components used to model the field are valid in a statistical sense and no attempt has been made to assign a physical interpretation or characterise the various components of the field (as done e.g for the cluster itself in the forthcoming sections).

The cluster is modelled with GMM in the astrometric space and a principal curve in the photometric space. The principal curve defines the cluster isochrone with a spread at any point that is given by a multivariate Gaussian. Both the principal curve and its spread are initialised with an input list of cluster members. The cluster model is built iteratively based on the initial list of cluster members that is provided to the algorithm in the first iteration of the process. The initial list of cluster members that we use consists of a sample of 88~stars compiled from the literature and located within the proper motion locus indicated in Figure~\ref{fig_pm_Lupus_lit}. This sample contains sources projected against the four main molecular clouds of the complex (Lupus 1 to 4) that are more likely to be associated with the Lupus region than to the Sco-Cen association (see Sect.~\ref{section2}). It contains 13~stars in Lupus~1, three~stars in Lupus~2, 47~stars in Lupus~3, nine~stars in Lupus~4, and 16~off-cloud stars. This first list of `candidate' members is indeed needed to start the membership analysis, but it can be incomplete and somewhat contaminated since it will be updated in the following iterations. Its main purpose is to locate the cluster locus in the space of astrometric observables and to define the empirical isochrone of the cluster in the photometric space to start the membership analysis. The algorithm uses the input list of members to infer the astrometric and photometric parameters of the cluster model. Then, the method assigns membership probabilities to the sources and classifies them into cluster and field stars based on a probability threshold $p_{in}$ previously defined by the user. At each step of the algorithm the marginal (i.e. data independent) class probabilities are estimated as the fraction of sources in each category obtained in the previous iteration. The resulting list of cluster members is used as input for the next iteration and this procedure is repeated until the list of cluster members remains fixed after successive iterations.  

Once that our solution has converged we generate synthetic data (based on the properties of the cluster and field models) and evaluate the performance of our classifier to define an optimum probability threshold $p_{opt}$ \citep[see][]{Olivares2019}. Finally, we classify the sources in our catalogue used for the membership analysis as members ($p\geq p_{opt}$) and non-members.  

Table~\ref{tab_comp_pin} shows the results of our membership analysis using different values for the user-defined probability threshold $p_{in}$. The true positive rate (TPR, i.e. fraction of cluster members generated from synthetic data that are recovered by the algorithm) and contamination rate (CR, i.e. fraction of field stars generated from synthetic data that are classified as cluster members by the algorithm) given in this table allow us to evaluate the quality of our membership analysis. However, we caution the reader that these figures derived from synthetic data are not absolute numbers but only estimates that are valid in the absence of the true distributions (and given the assumed cluster model). The solution obtained with $p_{in}=0.5$ exhibits the highest CR and contains fewer cluster members compared to the solutions derived with $p_{in}=0.6$ and $p_{in}=0.7$. The latter is explained from the more conservative $p_{opt}$ threshold that results due to the low initial threshold. On the other hand, the solutions derived from $p_{in}=0.8$ and $p_{in}=0.9$ have lower CRs (and higher TPRs), but a significantly reduced sample of cluster members. The solution obtained with $p_{in}=0.6$ offers the best compromise between the number of clusters members in the sample and the performance of our classifier (we note that the TPR and CR are compatible with the solutions obtained with the highest $p_{in}$ values). We have therefore adopted this solution (with 137 stars) as our final list of cluster members for the present study of the Lupus region. Table~\ref{tab_members} lists the 137 stars identified in our membership analysis together with other results that will be discussed in the forthcoming sections of this paper. We also provide in Table~\ref{tab_prob} the membership probabilities for all sources in the field using the $p_{in}$ values investigated in our analysis, so that the reader can select cluster members based on other criteria that are more specific to his/her scientific objectives. 

We tested the dependency of our membership analysis on the initial list of members that is provided to the algorithm in the first iteration. We selected the stars for the initial list using a somewhat narrower range of proper motions that concentrates most of the Lupus~3 candidate stars compiled from previous studies ($-13.0<\mu_{\alpha}\cos\delta< -8.0$ and $-25.0<\mu_{\delta}< -21.0$) and repeated the membership analysis as described before. We concluded based on similar arguments presented for the first run that the solution obtained with $p_{in}=0.6$ returns the best compromise between the sample size (141 members) and performance of our classifier ($TPR=0.91\pm0.02$ and $CR=0.13\pm0.04$) as compared to the other solutions with different $p_{in}$ thresholds. These numbers are consistent with the values obtained in the first run (see Table~\ref{tab_comp_pin}) and small variations are expected to occur given that our classifier exhibits a non-zero CR. We recovered 95\% of the stars obtained in the first run and confirm that our result for the membership analysis is not dependent on the initial list of stars that we use (as long as they define the general properties of the Lupus association).

\begin{figure*}[!h]
\begin{center}
\includegraphics[width=0.33\textwidth]{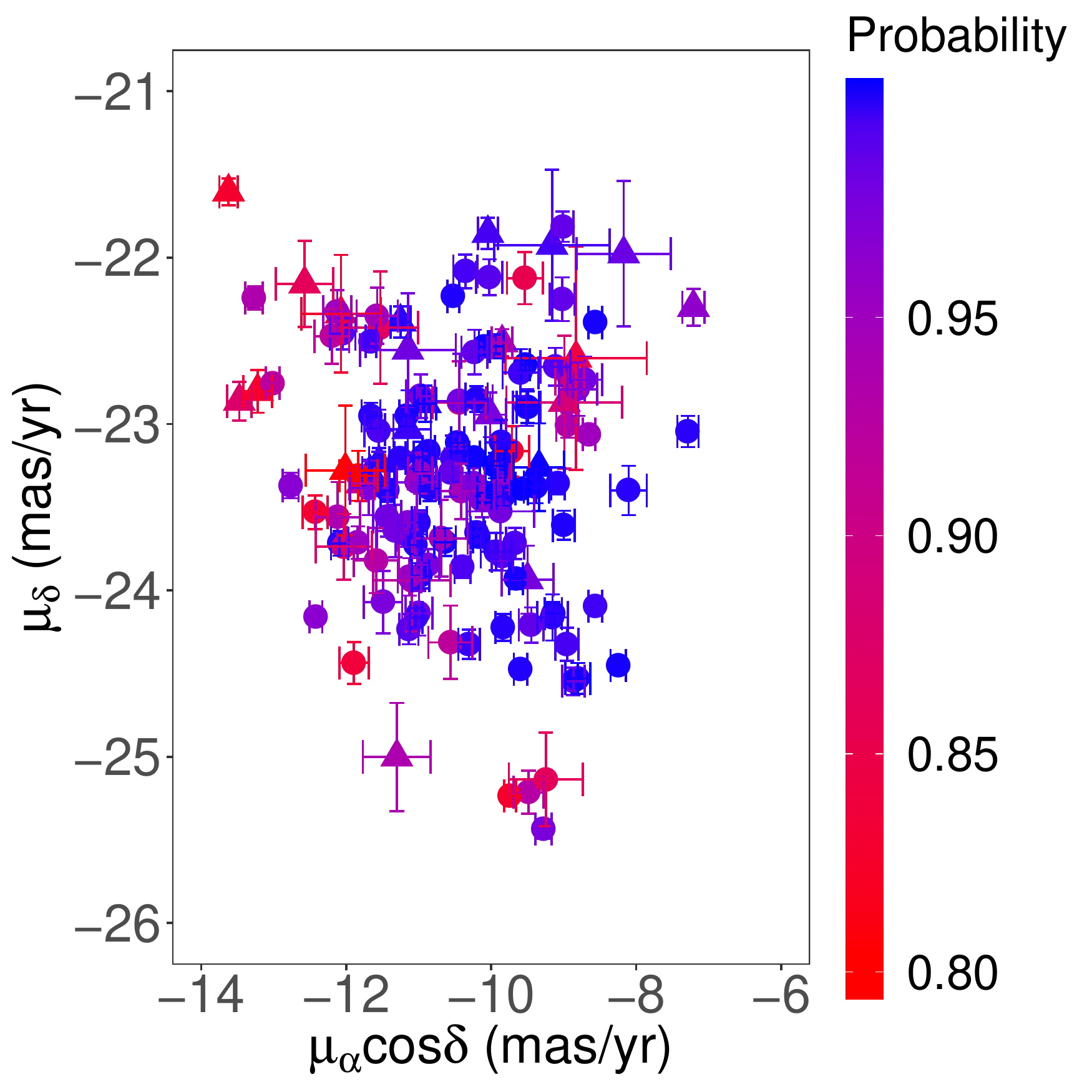}
\includegraphics[width=0.33\textwidth]{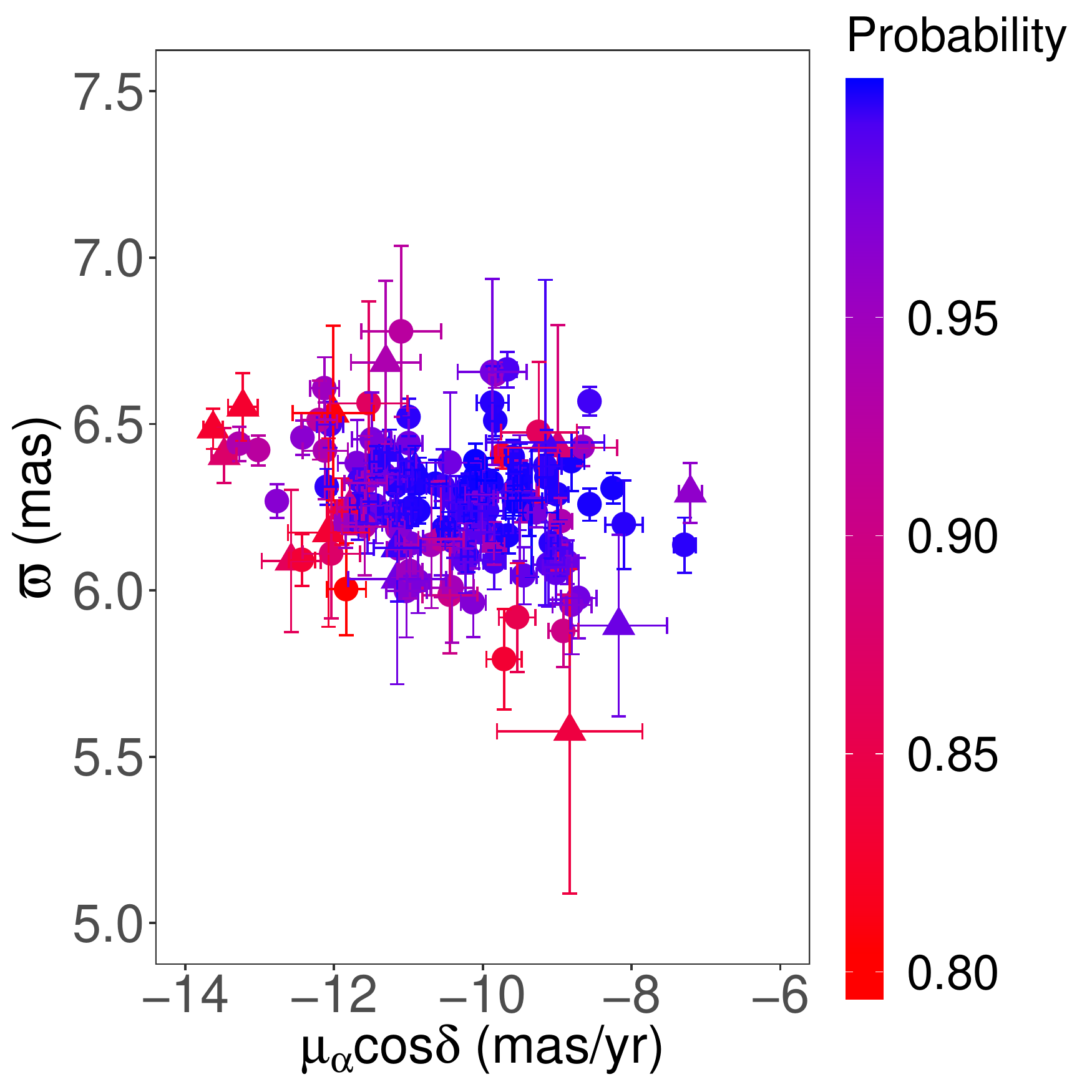}
\includegraphics[width=0.33\textwidth]{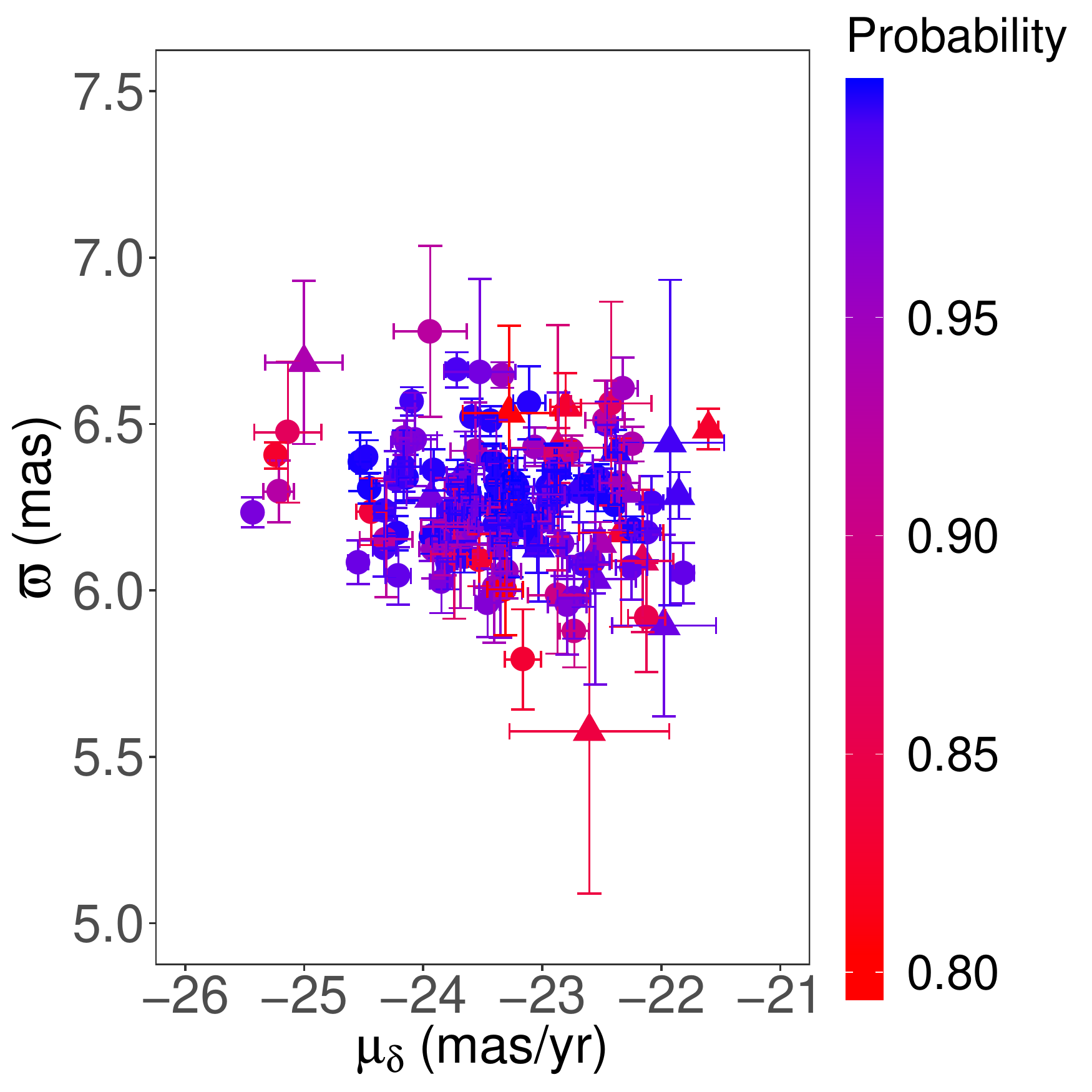}
\caption{Proper motions and parallaxes of the 137~Lupus stars identified in our membership analysis. The stars are colour-coded based on their membership probabilities which are scaled from zero to one.Triangles indicate the sources with RUWE $\geq$ 1.4 (see Sect.~\ref{section4.1}).
\label{fig_pm_plx_members} 
}
\end{center}
\end{figure*}

\begin{figure}[!h]
\begin{center}
\includegraphics[width=0.49\textwidth]{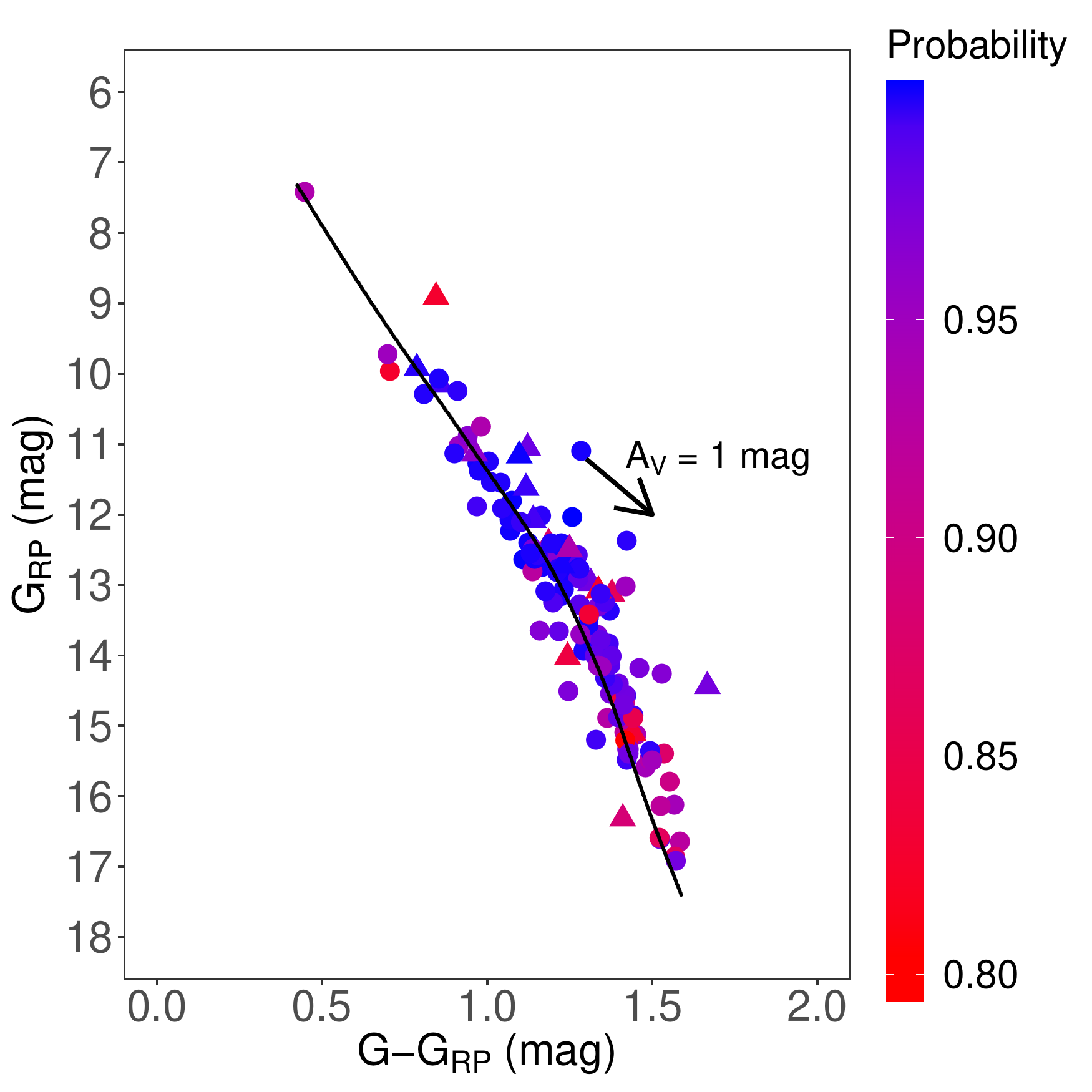}

\caption{Colour-magnitude diagram of the sample of Lupus stars identified in our membership analysis. The black solid line denotes the empirical isochrone of the Lupus association derived in this study (see Table~\ref{tab_isochrone}). The stars are colour-coded based on their membership probabilities which are scaled from zero to one. Triangles indicate the sources with RUWE $\geq$ 1.4 (see Sect.~\ref{section4.1}). The arrow indicates the extinction vector of $A_{V}=1$~mag converted to the \textit{Gai}a bands using the relative extinction values computed by \citet{Wang2019}.
\label{fig_CMD} 
}
\end{center}
\end{figure}

\begin{table}[!h]
\renewcommand\thetable{1} 
\centering
\caption{Comparison of membership results in the Lupus region using different values for the probability threshold $p_{in}$.
\label{tab_comp_pin}}
\begin{tabular}{ccccc}
\hline\hline
$p_{in}$&$p_{opt}$&Members&TPR&CR\\
\hline\hline
0.5&0.93 &113& $0.87 \pm 0.03$ & $0.18 \pm 0.08$\\
0.6&0.79 &137& $0.92 \pm 0.02$ & $0.10 \pm 0.04$\\ 
0.7&0.72 &133&$0.90 \pm 0.02$ & $0.12 \pm 0.03$\\ 
0.8&0.83 &102&$0.93 \pm 0.05$ & $0.09 \pm 0.03$\\ 
0.9&0.87 &73&$0.94\pm 0.02$ & $0.06 \pm 0.02$\\ 
\hline\hline
\end{tabular}
\tablefoot{We provide the optimum probability threshold, the number of cluster members, the true positive rate (TPR) and contamination rate (CR) obtained for each solution.}
\end{table}

\begin{figure*}[!h]
\begin{center}
\sidecaption
\includegraphics[width=0.9\textwidth]{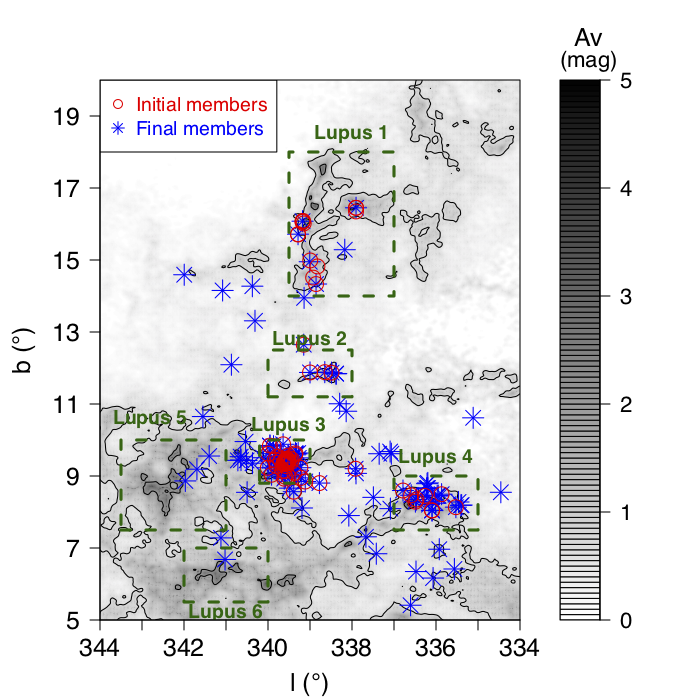}
\caption{Location of the Lupus stars overlaid on the extinction map of \citet{Dobashi2005} in Galactic coordinates. Colours and symbols denote the initial list of candidate stars compiled from the literature (88 stars) and the final members identified from our membership analysis (137 stars). The rectangles indicate the position of the main molecular clouds defined by \citet{Hara1999}.
\label{fig_location_Lupus_final} 
}
\end{center}
\end{figure*}

Figures~\ref{fig_pm_plx_members} and \ref{fig_CMD} show the proper motions, parallaxes and the colour-magnitude diagram of the Lupus stars selected in our membership analysis. The empirical isochrone that we derive from our membership analysis is given in Table~\ref{tab_isochrone}. Our sample of cluster members includes stars in the magnitude range from $G\simeq8$~mag to $G\simeq18$~mag. Most of the cluster members identified in our study are located in Lupus~3 and 4 as illustrated in Figure~\ref{fig_location_Lupus_final}. We note that Lupus~1 and 2 host only a minor fraction of cluster members. Interestingly, our sample includes three and one~stars projected towards Lupus~5 and 6, respectively. We confirm one star, namely Gaia~DR2~6021420630046381440, from \citet{Manara2018} and another three~stars (Gaia DR2 5995157042469914752, Gaia DR2 5997091667544824960 and Gaia DR2 5997086204346520448) from \citet{Melton2020} which are located in Lupus~5 and 6 as bona fide members of the region. We also confirm the existence of a more dispersed population of stars spread over the entire complex that is less numerous than the off-cloud population reported in pre-\textit{Gaia} studies \citep[see e.g.][]{Krautter1997,Wichmann1997b,Wichmann1997,Galli2013}. 

\begin{figure}[!h]
\begin{center}
\includegraphics[width=0.49\textwidth]{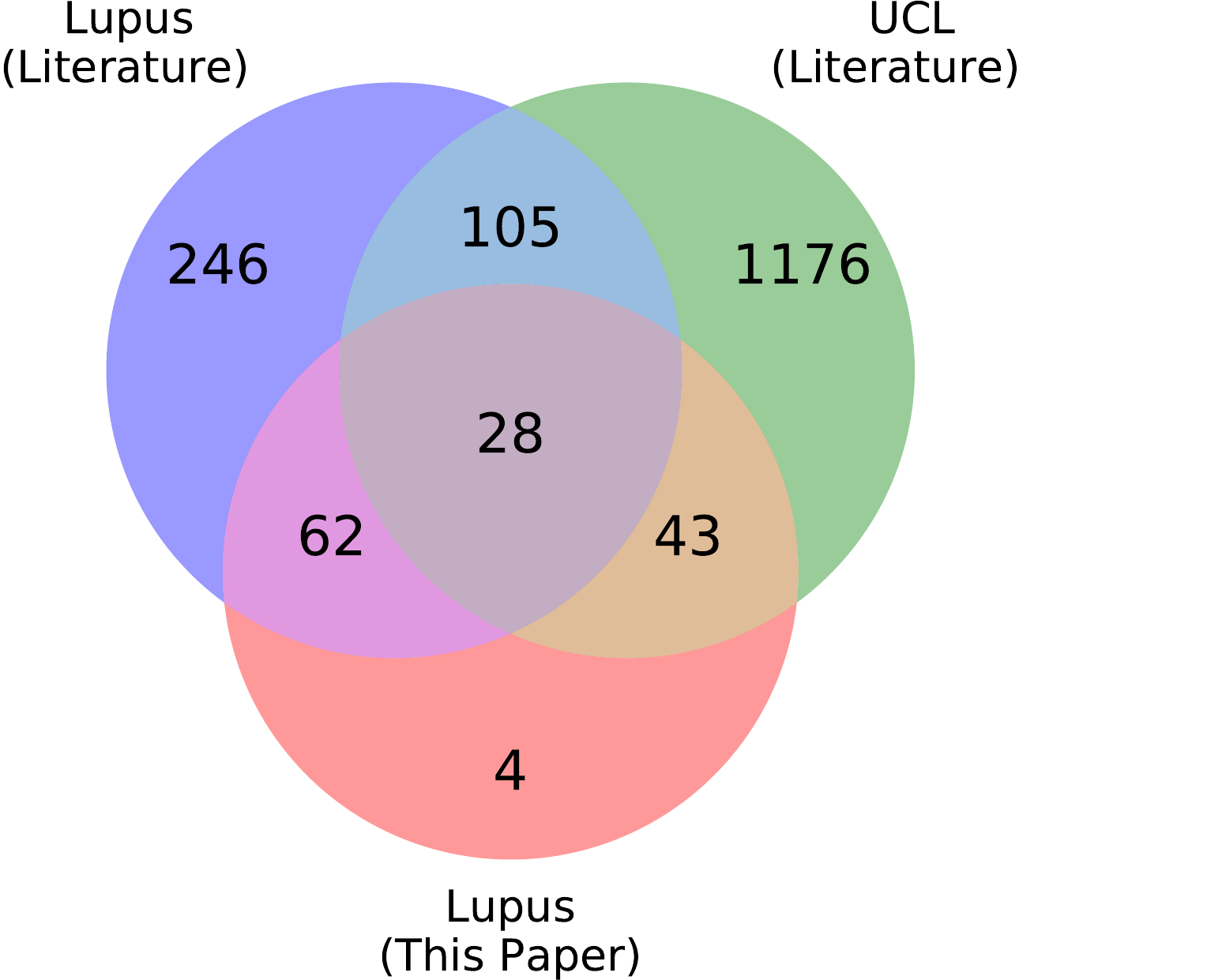}
\caption{Venn diagram comparing the number of Lupus stars compiled from the literature (with Gaia-DR2 data), Sco-Cen stars in the UCL subgroup, and the sample of confirmed Lupus members identified in this paper from our membership analysis.
\label{fig_venn} 
}
\end{center}
\end{figure}

The Venn diagram represented in Figure~\ref{fig_venn} shows that we have confirmed only 90 stars from previous studies as members of the Lupus association with our methodology. One important fraction of the literature sample with available data in Gaia-DR2, namely 351 stars, is rejected in our membership analysis. \citet{Damiani2019} classified 105 of them as Sco-Cen members. Among the remaining 246 sources we verify that most of them (i.e. 148 stars) have small parallaxes ($\varpi<3$~mas) confirming that they are indeed background stars unrelated to the Lupus region (as already anticipated in Sect.~\ref{section2}, see Figure~\ref{fig_pm_Lupus_lit}). We find more background sources at lower Galactic latitudes as also observed by \citet{Manara2018}. The remaining sample of 98 sources rejected from the literature includes 78 stars that  are mostly spread well beyond the location of the Lupus clouds in a region not covered by our survey ($l<334^{\circ}$) and exhibit different proper motions and parallaxes compared to the cluster members identified in this paper. The other 20 stars are indeed projected towards the molecular clouds of the complex (mostly in Lupus~3), but only one of them, namely Gaia DR2 5997013327325390592, has proper motion and parallax consistent with membership in Lupus. This star has $G_{RP}=11.77$~mag and $G-G_{RP}=1.65$~mag which puts it significantly above the empirical isochrone defined by the other cluster members (see e.g. Fig.~\ref{fig_CMD}). It was first proposed to be a member of the Lupus association by \citet{Damiani2019} and we found no more information about this star in the literature. We suspect that it is a binary or a high-order multiple system to explain its position in the colour-magnitude diagram.

A recent study conducted by \citet{Teixeira2020} and published during the revision process of this paper reports on the discovery of new YSOs with circumstellar discs in the Lupus complex. Two points are worth mentioning regarding the comparison of our sample of cluster members with their results. First, the survey conducted by \citet{Teixeira2020} covers a larger area in the sky than our study and extends clearly beyond the location of the Lupus molecular clouds. They identified 60 new YSOs that are potential members of the Lupus complex (these sources are labelled as `Lupus and/or UCL' in Table~4 of that study), but only 39 of them are included in the field selected for our membership analysis in this paper. Second, the methodology used in the two studies to select cluster members differs significantly. \citet{Teixeira2020} selected their sample of cluster members from proper motion cuts based on Lupus stars previously identified in the literature. As shown in Figure~\ref{fig_venn}, many of the Lupus stars identified in the literature are probable field stars (unrelated to the Lupus clouds) and the adopted proper motion constraints in that study to select new members (i.e. $-27<\mu_{\alpha}\cos\delta<3$~mas/yr and $-33<\mu_{\delta}<-13$~mas/yr) extend to a region in the proper motion vector diagram that is mostly populated by UCL stars (see Figure~\ref{fig_pm_Lupus_lit}). As a result, we found only 3 stars (out of the sample of 39 sources) that are in common with the study of \citet{Teixeira2020}. The remaining 36 sources listed by \citet{Teixeira2020} and included in our analysis have mostly small membership probabilities suggesting that they are unrelated to Lupus. Indeed, the authors themselves concluded in their study that the older stars in the sample are most probably UCL members. 

In this study, we report on the discovery of 47 new members of the Lupus association, which represents an increase of more than 50\% with respect to the number of confirmed members already known from previous studies. Most of them (i.e. 43 stars) had previously been assigned to the Sco-Cen association by \citet{Damiani2019}. One reason to explain this is that the authors included only Lupus~3 in their membership analysis so that the Lupus stars belonging to the other subgroups were either assigned to Sco-Cen or to the field population. The 43 stars populate all subgroups of the Lupus complex (except for Lupus~3) and our membership analysis focused on the properties of the whole region was able to identify them as members. In the following we use our new sample of 137 members to revisit some properties of the Lupus star-forming region.

\newpage
\section{Properties of the Lupus subgroups}\label{section4}

\subsection{Refining the sample of Lupus stars}\label{section4.1}

We refine our list of cluster members by selecting the stars with the best astrometry in the sample to accurately determine the properties of the Lupus subgroups. We remove 24~stars with poor astrometric solution based the re-normalised unit weight error (RUWE) criterion\footnote{see technical note \href{https://www.cosmos.esa.int/web/gaia/dr2-known-issues}{GAIA-C3-TN-LU-LL-124-01} for more details.} (i.e. with RUWE $\geq$ 1.4). These sources exhibit the largest uncertainties in proper motions and parallaxes (see Fig.~\ref{fig_pm_plx_members}) so that their membership status will require further investigation with future data releases of the \textit{Gaia} space mission. This step reduces the sample of cluster members to 113~stars.

\subsection{Proper motions and parallaxes}

\begin{figure*}[!h]
\begin{center}
\includegraphics[width=0.33\textwidth]{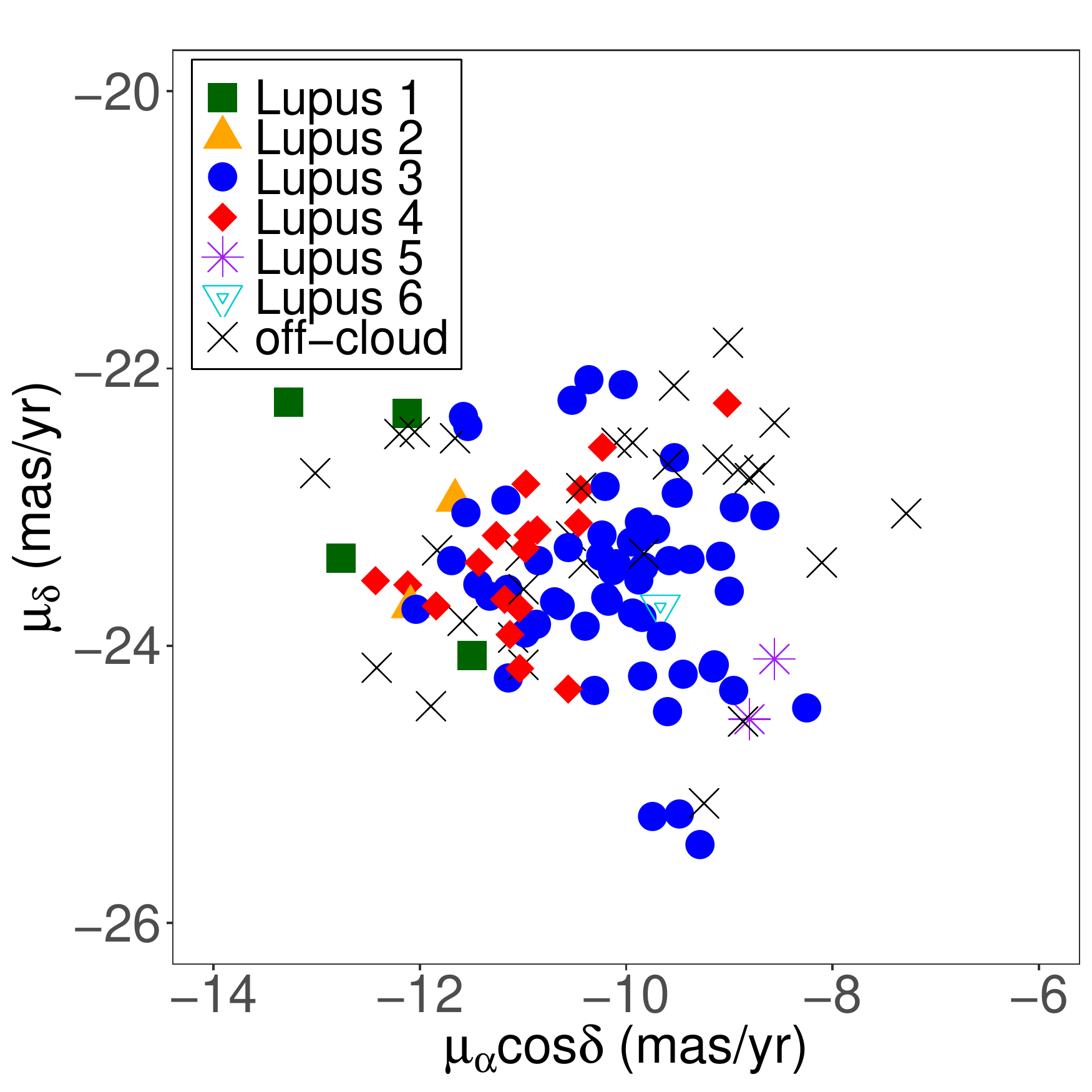}
\includegraphics[width=0.33\textwidth]{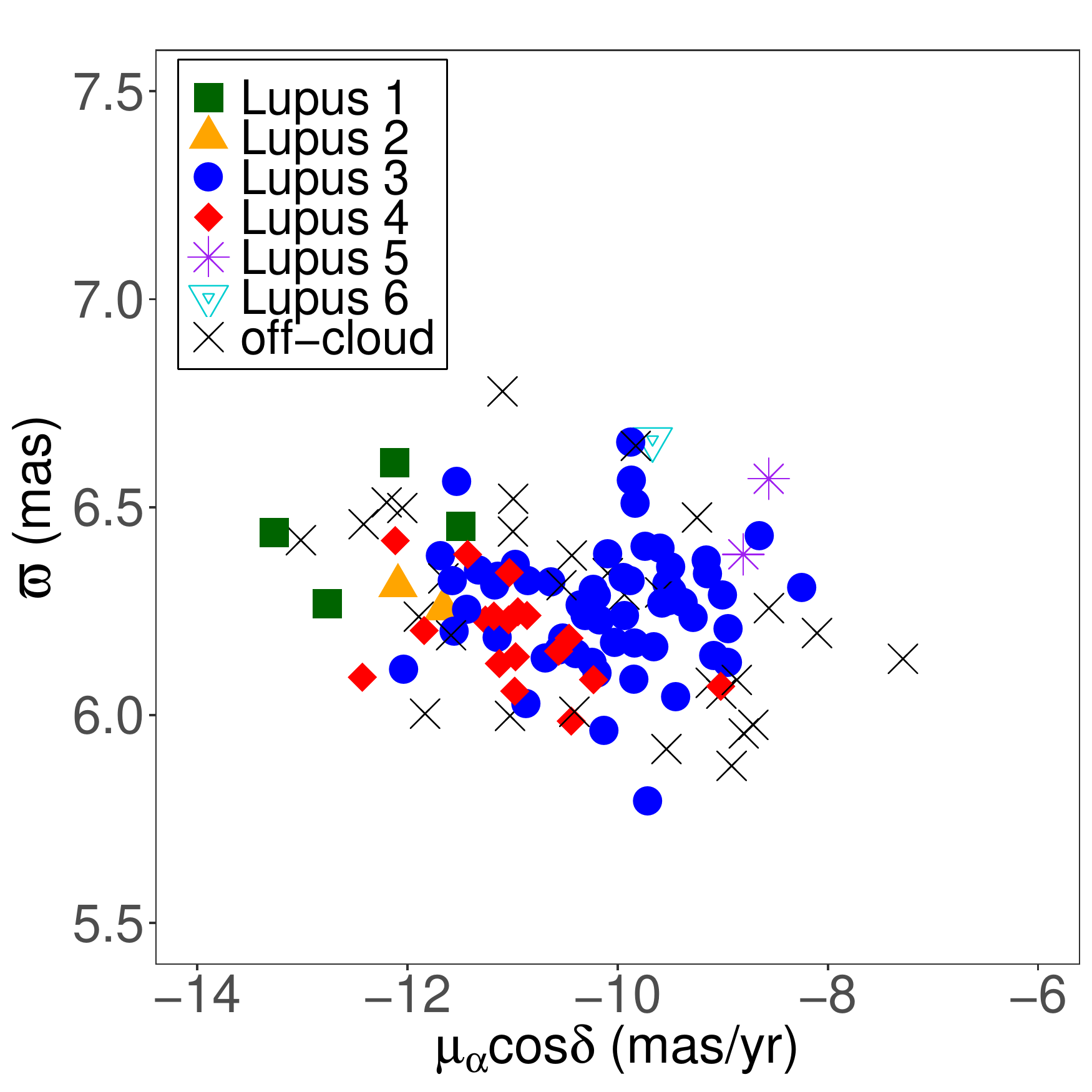}
\includegraphics[width=0.33\textwidth]{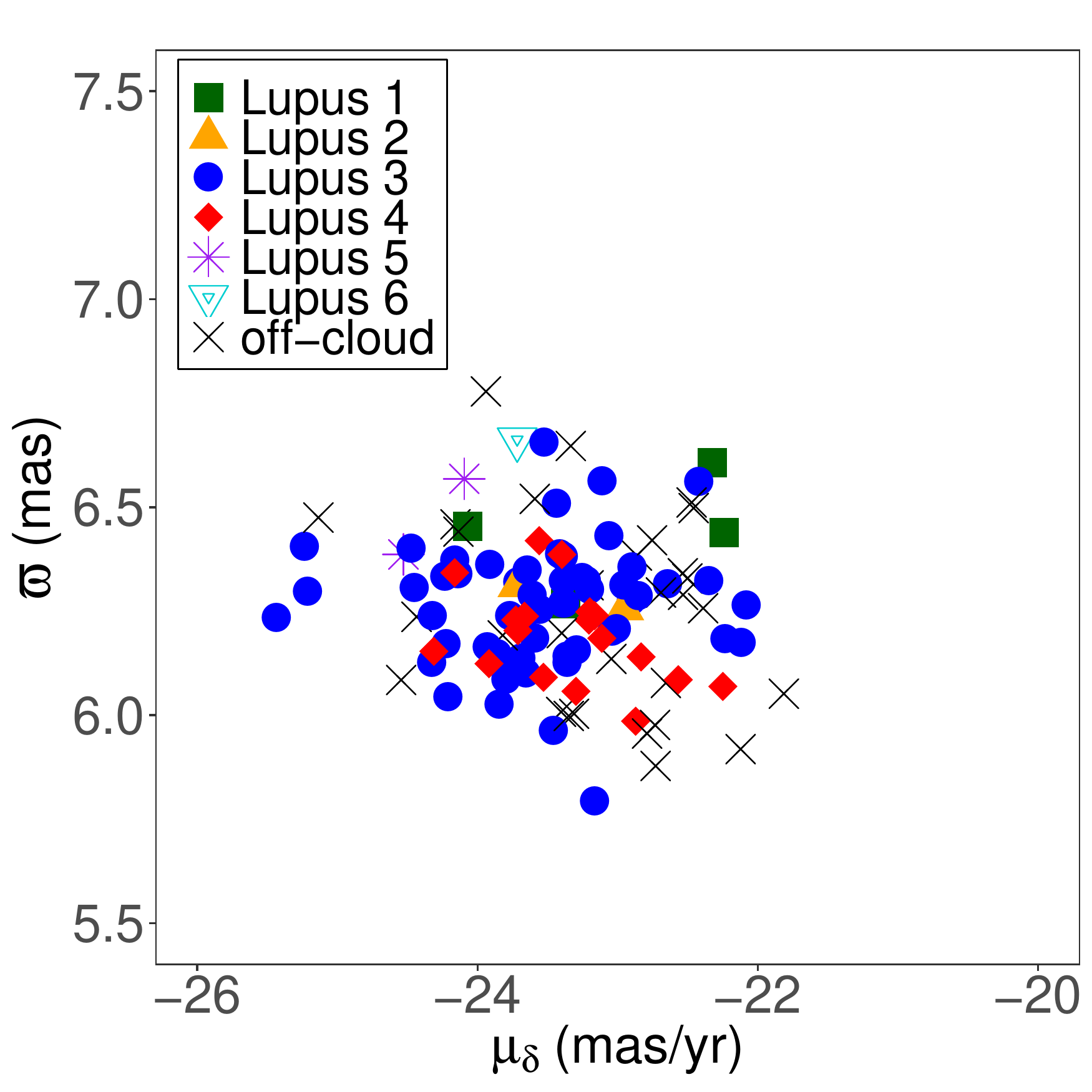}
\caption{Proper motions and parallaxes for the Lupus subgroups in our sample of cluster members. 
\label{fig_pm_plx_subgroups} 
}
\end{center}
\end{figure*}

Figure~\ref{fig_pm_plx_subgroups} shows the distribution of proper motions and parallaxes for the several subgroups of the Lupus complex. The mean values of each parameter are listed in Table~\ref{tab_subgroups}. A first inspection of the sample reveals that the stars projected towards Lupus~3 exhibit different proper motions (in right ascension) that are shifted by about 1-2~mas/yr with respect to the stars in Lupus~1, 2 and 4. This proper motion shift translates into 0.8-1.5~km/s in the tangential velocities of the subgroups. On the other hand, we see no important differences between the proper motion components in declination and parallaxes among the various subgroups of the Lupus complex. 

We performed a two sample Kolmogorov-Smirnov test to quantitatively confirm our findings. We compare the distribution of proper motions and parallaxes from Lupus~3 stars with respect to the stars in the other molecular clouds (Lupus~1, 2, 4). We find the $p$-values of 1.97$\times$10$^{-6}$, 0.64, and 0.23, for the analyses based on $\mu_{\alpha}\cos\delta$, $\mu_{\delta}$ and $\varpi$, respectively. Thus, we conclude that the proper motion component in right ascension is the most distinctive astrometric feature that distinguishes the stars projected towards Lupus~3 from the other subgroups in the region. The few stars in Lupus~5 and 6 that have been retained in our sample have proper motions that are more consistent with Lupus~3 stars. 

This small difference in proper motions among the subgroups is seen now for the first time with the more precise Gaia-DR2 data that is available nowadays.  Previous studies about the kinematic properties of the Lupus region \citep[see e.g.][]{Makarov2007,Lopez-Marti2011,Galli2013} used proper motions measured from the ground with a typical precision of about 2~mas/yr that was not enough to detect subtle differences of the stars located in different subgroups. The median uncertainties in proper motions and parallaxes for the Lupus stars in our sample based on Gaia-DR2 data are $\sigma_{\mu_{\alpha}\cos\delta}=0.15$~mas/yr, $\sigma_{\mu_{\delta}}=0.09$~mas/yr and $\sigma_{\varpi}=0.08$~mas, respectively. This represents an improvement in precision of about one order of magnitude with respect to pre-\textit{Gaia} studies. 

\begin{table*}[!h]
\centering
\scriptsize{
\caption{Proper motions and parallaxes of the Lupus subgroups in our sample of cluster members.
\label{tab_subgroups}}
\begin{tabular}{lccccccccccc}
\hline\hline
Sample&$N_{init}$&$N_{RUWE}$&\multicolumn{3}{c}{$\mu_{\alpha}\cos\delta$}&\multicolumn{3}{c}{$\mu_{\delta}$}&\multicolumn{3}{c}{$\varpi$}\\
&&&\multicolumn{3}{c}{(mas/yr)}&\multicolumn{3}{c}{(mas/yr)}&\multicolumn{3}{c}{(mas)}\\
\hline\hline
&&&Mean$\pm$SEM&Median&SD&Mean$\pm$SEM&Median&SD&Mean$\pm$SEM&Median&SD\\
\hline

Lupus 1 & 6 & 4 & $ -12.414 \pm 0.387 $& -12.446 & 0.773 & $ -23.002 \pm 0.439 $& -22.848 & 0.878 & $ 6.442 \pm 0.069 $& 6.447 & 0.139 \\
Lupus 2 & 5 & 2 & $ -11.875 \pm 0.216 $& -11.875 & 0.433 & $ -23.335 \pm 0.383 $& -23.335 & 0.767 & $ 6.283 \pm 0.028 $& 6.283 & 0.056 \\
Lupus 3 & 61 & 56 & $ -10.118 \pm 0.113 $& -9.986 & 0.843 & $ -23.541 \pm 0.095 $& -23.492 & 0.710 & $ 6.262 \pm 0.020 $& 6.272 & 0.148 \\
Lupus 4 & 22 & 18 & $ -10.997 \pm 0.177 $& -11.006 & 0.753 & $ -23.361 \pm 0.125 $& -23.348 & 0.532 & $ 6.190 \pm 0.027 $& 6.194 & 0.116 \\
Lupus 5 & 3 & 2 & $ -8.684 \pm 0.121 $& -8.684 & 0.241 & $ -24.313 \pm 0.217 $& -24.313 & 0.434 & $ 6.477 \pm 0.091 $& 6.477 & 0.182 \\
Lupus 6 & 1 & 1 & $ -9.669 \pm 0.125 $& -9.669 & \nodata & $ -23.718 \pm 0.094 $& -23.718 & \nodata & $ 6.663 \pm 0.053 $& 6.663 & \nodata \\
Lupus (off-cloud) & 39 & 30 & $ -10.259 \pm 0.263 $& -10.254 & 1.443 & $ -23.162 \pm 0.143 $& -22.954 & 0.782 & $ 6.256 \pm 0.042 $& 6.273 & 0.231 \\
\hline
Lupus (all stars) & 137 & 113 & $ -10.378 \pm 0.109 $& -10.306 & 1.155 & $ -23.404 \pm 0.068 $& -23.387 & 0.721 & $ 6.263 \pm 0.017 $& 6.265 & 0.178 \\

\hline\hline
\end{tabular}
\tablefoot{We provide for each subgroup the initial number of stars, final number of stars after the RUWE filtering (see Sect.~\ref{section4.1}), mean, standard error of the mean (SEM), median, and standard deviation (SD) of proper motions and parallaxes.}
}
\end{table*}

\subsection{Radial velocities}

We search for radial velocity information for the stars in our sample using the CDS databases and data mining tools \citep{Wenger2000}. Our search in the literature is based on \citet{Wichmann1999}, \citet{Gontcharov2006}, \citet{Torres2006}, \citet{James2006}, \citet{Guenther2007}, \citet{Galli2013}, \citet{Frasca2017}, and the Gaia-DR2 catalogue. We find radial velocities for 52~stars in the sample of 113~sources with good astrometry (see Sect.~\ref{section4.1}). For a few stars we find more than one single radial velocity measurement and in such cases we use the most precise result. 

\begin{figure}[!h]
\begin{center}
\includegraphics[width=0.49\textwidth]{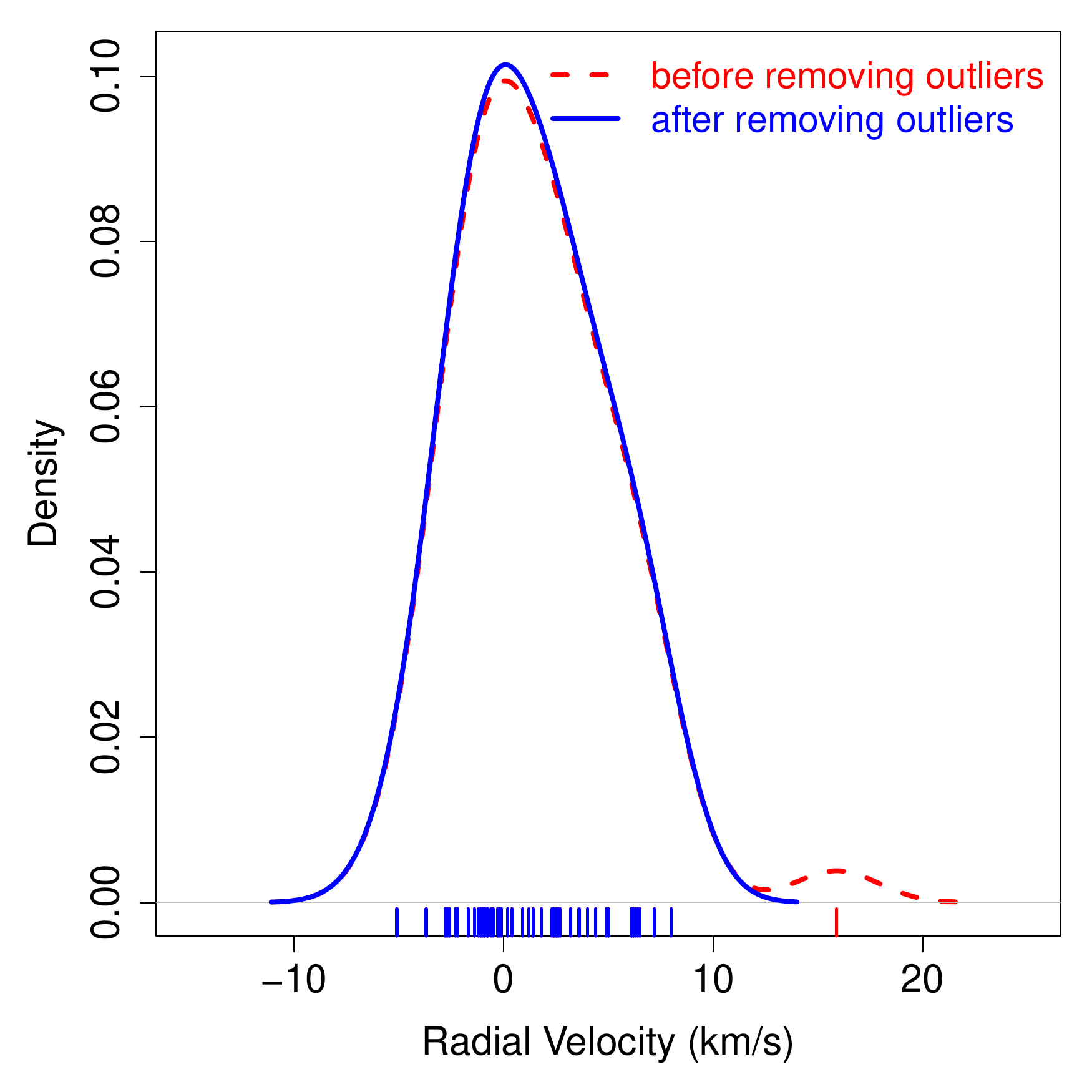}
\caption{Kernel density estimate of the distribution of radial velocity measurements for the sample of 52 members with available data. We used a kernel bandwidth of 2~km/s. The tick marks in the horizontal axis mark the individual radial velocity values of each star.
\label{fig_rv} 
}
\end{center}
\end{figure}

The radial velocity distribution of the Lupus stars in our sample is shown in Figure~\ref{fig_rv}. We use the interquartile range (IQR) criterium to identify the outliers that lie over 1.5$\times$IQR below the first quartile or above the third quartile of the distribution. Only the radial velocity of Gaia~DR2~5997015079686716672 satisfies this condition. We discard this radial velocity measurement \citep[$V_{r}=15.9\pm0.7$~km/s,][]{Frasca2017}, but we still retain the star in the sample because its proper motion and parallax are consistent with membership in Lupus. 

We find that 20~candidate members in our sample of 113~stars have been identified as binaries or multiple systems in previous studies \citep{Ghez1997,Merin2008}. Nineteen of them have radial velocity information in the literature, but only EX~Lup (Gaia DR2 5996902860784332800) shows periodic radial velocity variations most probably due to a substellar companion \citep{Kospal2014}. Analogously, we retain this star in the sample, but we do not use its radial velocity in the forthcoming analysis. Altogether, this reduces the sample of sources with radial velocity information to 50~stars. 

Table~\ref{tab_subgroups_RV} lists the mean radial velocity values for the Lupus subgroups (except for Lupus~5 and 6 which have no published data). We find no significant difference between the mean radial velocities of the subgroups within the uncertainties. The admittedly large scatter of radial velocities that is reported here is most probably related to the precision of the individual radial velocity measurements. The median uncertainty of the radial velocities in our sample is 2.2~km/s. Removing the binaries from the sample has negligible impact in reducing the radial velocity scatter. On the other hand, we note that the median uncertainty on the tangential velocities derived from Gaia-DR2 data is only 0.3~km/s and 0.4~km/s in the right ascension and declination components, respectively. Thus, the precision of the radial velocity measurements reported in the literature is currently the main limitation to investigate the dynamics (e.g. internal motions, expansion and rotation effects) of the Lupus complex. 

\begin{table}[!h]
\centering
\scriptsize{
\caption{Radial velocity of the Lupus subgroups in our sample.
\label{tab_subgroups_RV}}
\begin{tabular}{lccccc}
\hline\hline
Sample&$N$&\multicolumn{4}{c}{$V_{r}$}\\
&&\multicolumn{4}{c}{(km/s)}\\
\hline\hline
&&Mean$\pm$SEM&Median&SD&$\sigma_{V_{r}}$\\
\hline

Lupus 1 & 2 & $ -0.7 \pm 2.2 $& -0.7 & 4.1 & 1.5\\
Lupus 2 & 2 & $ -0.6 \pm 3.1 $& -0.6 & 0.9 & 1.9\\
Lupus 3 & 35 & $ 1.5 \pm 0.6 $& 1.2 & 3.5 & 2.2\\
Lupus 4 & 7 & $ 2.1 \pm 0.7 $& 2.5 & 1.9 & 2.2\\
Lupus (off-cloud) & 4 & $ 1.0 \pm 2.1 $& 0.2 & 4.3 &1.8 \\
\hline
Lupus (all stars) & 50 & $ 1.4 \pm 0.5 $& 1.3 & 3.2 & 2.2\\

\hline\hline
\end{tabular}
\tablefoot{We provide for each subgroup the initial number of stars, mean, standard error of the mean (SEM), median, standard deviation (SD) of the radial velocities, and the median uncertainty of the radial velocities.}
}
\end{table}

\subsection{Distance and spatial velocity}\label{sect4.4}

In this section we re-visit the distance and kinematic properties of the Lupus region based on our new sample of cluster members, the more precise Gaia-DR2 astrometry and the radial velocity measurements collected from the literature. The first step consists in correcting the Gaia-DR2 parallaxes by the zero-point shift of -0.030~mas that is present in the published data \citep[see e.g.][]{Lindegren2018}. There are different estimates of the zero-point correction for the Gaia-DR2 parallaxes in the literature that range from $-0.031\pm0.011$~mas \citep{Graczyk2019} to $-0.082\pm0.033$~mas \citep{Stassun2018}, but we verified that the impact of this correction in our solution is negligible given the close proximity of the Lupus clouds. In addition, we also add 0.1~mas/yr in quadrature to the formal uncertainties on the proper motions of the stars to take the systematics errors of the Gaia-DR2 catalogue into account \citep[see e.g.][]{Lindegren2018}. This procedure is likely to overestimate the proper motion uncertainty for some stars in the sample given that the systematic errors in the Gaia-DR2 catalogue depend on the position, magnitude and colour of each source \citep[see e.g.][]{Luri2018}. These corrections are necessary for the purpose of estimating distances and spatial velocities with the corresponding uncertainties taking into account systematic errors, but they do not affect our results of the membership analysis. 

Then, we used Bayesian inference to convert the observed parallaxes and proper motions into distances and 2D tangential velocities. The prior that we used for the angular velocity (i.e. 2D tangential velocity) is a beta function following the online tutorials available in the \textit{Gaia} archive \citep[see e.g.][]{Luri2018}. We investigated two types of prior families for the distance. The first one is based on purely statistical probability density distributions (hereafter, statistical priors) and the second type is inspired on astrophysical assumptions (hereafter, astrophysical priors) e.g. the luminosity and density profiles of stellar clusters. The statistical priors investigated in this study are based on the Uniform and Gaussian distributions. The astrophysical priors include the surface brightness profile derived from star counts in the Large Magellanic Cloud by \citet[][hereafter, the EFF prior]{Elson1987} and the King's profile distribution derived from globular clusters \citep[][hereafter, the King prior]{King1962}. We used the \textit{Kalkayotl} code \citep[][Olivares et al. 2020, A\&A in press\footnote{The code is available at \href{https://github.com/olivares-j/Kalkayotl}{https://github.com/olivares-j/Kalkayotl}.}]{kalkayotl} to compute the distances with these different prior families. We refer the reader to that paper for more details on the properties, validation and performance of these priors. We are aware about the existence of another prior in the literature, namely the exponentially decreasing space density prior \citep[hereafter, the EDSD prior,][]{Bailer-Jones2015,Astraatmadja2016}, but we do not use it in our study. The EDSD prior is related to the distribution of stars in the Galaxy and it was proposed to be used in the context of large samples with wide distributions of parallaxes and uncertainties which is not the case of the Lupus sample.

We find no significant differences in the resulting distances obtained with different priors (Uniform, Gaussian, EFF and King) which agree within 1$\sigma$ of the corresponding distance uncertainties. Thus, we confirm that our results are not affected by our choice of the prior. We have therefore decided to use the distances obtained with the Uniform prior, which is the simplest one, and report them as our final results. The distance and spatial velocity for individual stars in our sample are given in Table~\ref{tab_members} together with other results derived in this paper. We use the resulting distances to compute the 3D position $XYZ$ of the stars in a right-handed reference system that has its origin at the Sun, where $X$ points to the Galactic centre, $Y$ points in the direction of Galactic rotation, and $Z$ points to the Galactic north pole. Then, we convert the 2D tangential velocity and radial velocity of the stars into the $UVW$ Galactic velocity components in this same reference system following the transformation outlined by \citet{Johnson1987}. 

\begin{figure*}[!h]
\begin{center}
\sidecaption
\includegraphics[width=0.95\textwidth]{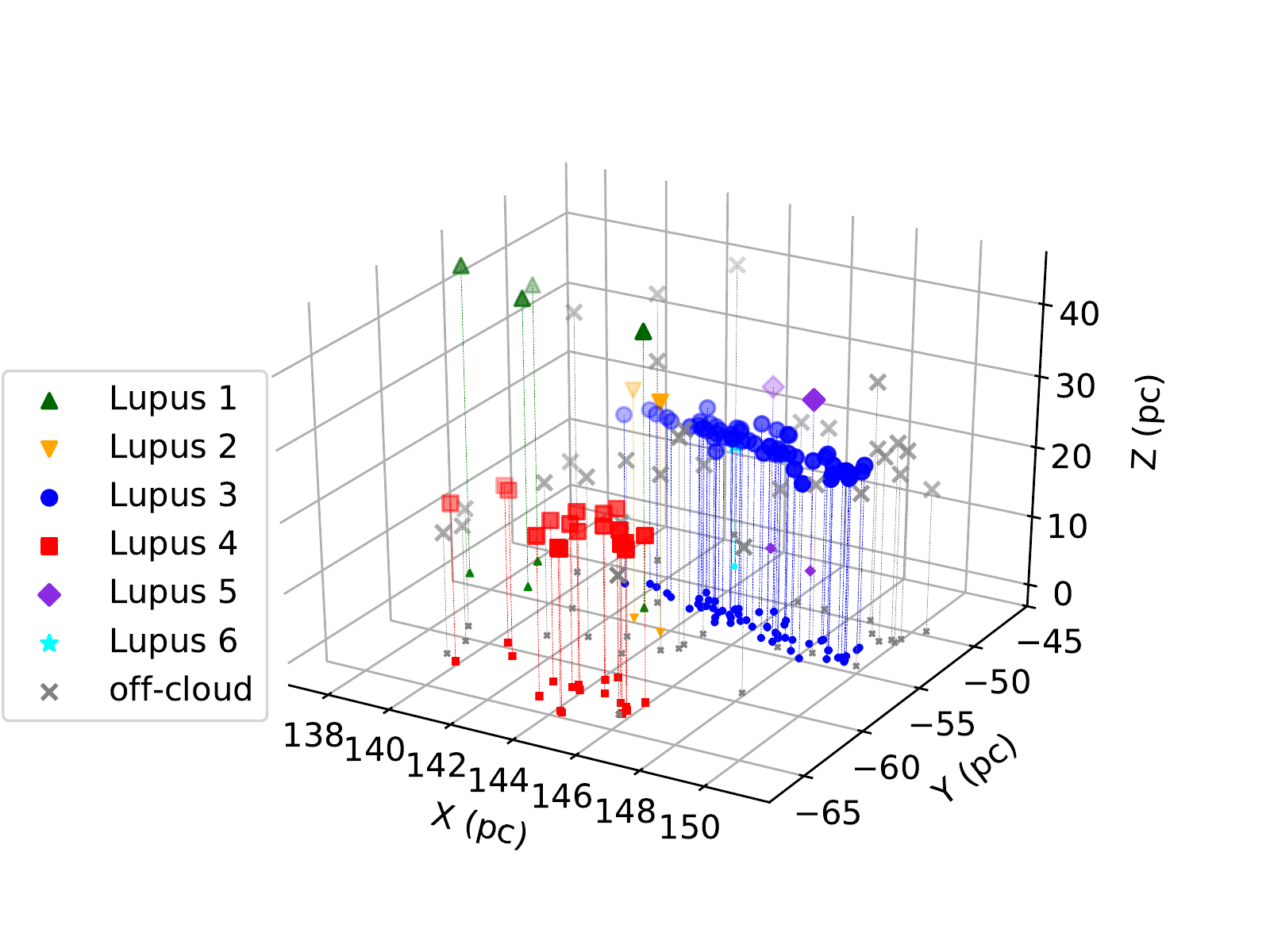}
\caption{3D spatial distribution of the 113 members of the Lupus association. The different colours denote the various subgroups of the Lupus star-forming region. The dashed lines connect each point in the plot to its projection on the XY plane (with Z=0). The reference system used to represent the 3D spatial distribution of the stars is the same as described in the text of Sect.~\ref{sect4.4}.
\label{fig_3d} 
}
\end{center}
\end{figure*}

Table~\ref{tab_distance_velocity} lists the distance and spatial velocity of the Lupus subgroups in our sample. The large uncertainties of our distance results observed for some subgroups can be explained by the small number of stars used in each solution. Our results reveal that the Lupus subgroups are located at the same distance (of about 160~pc) within the uncertainties derived for each solution. The only exception is Lupus~6 whose distance estimate given in Table~\ref{tab_distance_velocity} is based on the parallax of one single star. Despite the common distance to the Sun, the subgroups (Lupus 1 to 4) are separated by about 8-20~pc among themselves in the space of 3D positions as illustrated in Figure~\ref{fig_3d}.  

The distances derived in this paper are still consistent with the results obtained by \citet{Dzib2018} despite the different samples of stars that have been used in each study. Our results do not confirm the large spread in distances of tens of parsec along the line of sight reported in previous studies based on kinematic parallaxes \citep{Makarov2007,Galli2013} and \textit{Hipparcos} data \citep[see e.g.][]{Bertout1999,Comeron2008}. One reason to explain the discrepancy with previous findings in the pre-\textit{Gaia} era is the more precise and accurate data that is available nowadays to investigate the Lupus region. Another reason is the different samples of Lupus stars used in each study. For example, \citet{Makarov2007} and \citet{Galli2013} included the more dispersed stars identified by \citet{Krautter1997} and \citet{Wichmann1997b,Wichmann1997} in their solutions (see Fig.~\ref{fig_location_Lupus_lit}) which are mostly rejected in our membership analysis. In addition, many of the stars projected towards the molecular clouds and included in their analyses appear now to be background sources unrelated to Lupus in light of the Gaia-DR2 data (see Sect.~\ref{section3}). 

The mean spatial velocities of the Lupus subgroups given in  Table~\ref{tab_distance_velocity} confirm that they move at a common speed. We therefore detect no significant relative motion between the subgroups at the level of a few km/s which is currently determined by the precision of radial velocity data. The off-cloud population is located at the same distance and moves with the same velocity of the stars projected towards Lupus~1 to 4. This confirms that they are comoving with the on-cloud population and therefore belong to the same association. 

We now compare the distance and kinematics of Lupus with the two adjacent subgroups of the Sco-Cen association (US and UCL). The relative motion of Lupus with respect to the space motion of US and UCL derived by \citet{Wright2018} is $(\Delta U,\Delta V,\Delta W)=(2.4,-0.7,-0.4)\pm(0.4,0.2,0.1)$~km/s and $(\Delta U,\Delta V,\Delta W)=(2.1,2.4,-1.6)\pm(0.4,0.2,0.1)$~km/s, respectively (in the sense Lupus `minus' US or UCL). This implies a relative bulk motion of $2.5\pm0.4$
~km/s and $3.6\pm0.3$~km/s with respect to US and UCL, respectively. \citet{Damiani2019} revealed that Sco-Cen is made up of several substructures and the distances of the stars range from about 100 to 200~pc. The closest substructure of the Sco-Cen association to the Lupus molecular cloud complex is a compact group of 551 stars near V1062~Sco and named as UCL-1 \citep{Roeser2018,Damiani2019}. The mean parallax of $\varpi=5.668\pm0.011$~mas and distance of $174.7^{+0.5}_{-0.6}$~pc put UCL-1 further away from the Sun as compared to Lupus (see Table~\ref{tab_distance_velocity}). Thus, despite the common location in the plane of the sky Lupus and Sco-Cen stars are located at different distances and exhibit distinct spatial velocities. Interestingly, we note that the observed difference between the spatial velocity of the Lupus clouds, UCL and US is comparable to the difference in spatial velocity of the Sco-Cen subgroups among themselves. The Lupus clouds could therefore be regarded as another subgroup of the Sco-Cen complex that resulted from a more recent star formation episode.

\begin{table*}[!h]
\centering
\scriptsize{
\caption{Distance and spatial velocity of the Lupus subgroups in our sample of cluster members.
\label{tab_distance_velocity}}
\begin{tabular}{lcccccccccccc}
\hline\hline
Sample&$N_{d}$&$N_{UVW}$&$d$&\multicolumn{3}{c}{$U$}&\multicolumn{3}{c}{$V$}&\multicolumn{3}{c}{$W$}\\
&&&(pc)&\multicolumn{3}{c}{(km/s)}&\multicolumn{3}{c}{(km/s)}&\multicolumn{3}{c}{(km/s)}\\
\hline\hline
&&&&Mean$\pm$SEM&Median&SD&Mean$\pm$SEM&Median&SD&Mean$\pm$SEM&Median&SD\\
\hline

Lupus 1 & 4 & 2 & $ 154.9 _{ -3.4 }^{+ 3.2 }$ &$ -5.5 \pm 1.8 $& -5.5 & 2.5 & $ -17.5 \pm 1.2 $& -17.5 & 1.7 & $ -6.9 \pm 0.2 $& -6.9 & 0.3 \\
Lupus 2 & 2 & 2 & $ 157.7 _{ -5.3 }^{+ 6.9 }$ &$ -5.8 \pm 0.5 $& -5.8 & 0.7 & $ -17.4 \pm 0.1 $& -17.4 & 0.1 & $ -7.3 \pm 0.1 $& -7.3 & 0.2 \\
Lupus 3 & 56 & 35 & $ 158.9 _{ -0.7 }^{+ 0.7 }$ &$ -3.5 \pm 0.5 $& -3.6 & 3.1 & $ -17.6 \pm 0.3 $& -17.5 & 1.5 & $ -7.5 \pm 0.2 $& -7.5 & 0.9 \\
Lupus 4 & 18 & 7 & $ 160.2 _{ -0.9 }^{+ 0.9 }$ &$ -4.3 \pm 0.6 $& -4.3 & 1.7 & $ -17.9 \pm 0.4 $& -18.2 & 1.0 & $ -7.3 \pm 0.2 $& -7.1 & 0.5 \\
Lupus 5 & 2 & 0 & $ 155.5 _{ -7.4 }^{+ 13.4 }$ &\nodata& \nodata & \nodata & \nodata& \nodata&\nodata & \nodata&\nodata & \nodata \\
Lupus 6 & 1 & 0 & $ 149.4 _{ -1.9 }^{+ 1.7 }$ &\nodata& \nodata & \nodata & \nodata&\nodata & \nodata & \nodata&\nodata & \nodata \\
Lupus (off-cloud) & 30 & 4 & $ 158.9 _{ -1.0 }^{+ 1.0 }$ &$ -3.7 \pm 2.0 $& -4.4 & 3.9 & $ -17.2 \pm 0.8 $& -17.4 & 1.6 & $ -7.1 \pm 0.4 $& -7.3 & 0.9 \\
\hline
Lupus (all stars) & 113 & 50 & $ 158.3 _{ -0.6 }^{+ 0.6 }$ &$ -3.8 \pm 0.4 $& -4.4 & 2.9 & $ -17.6 \pm 0.2 $& -17.5 & 1.4 & $ -7.4 \pm 0.1 $& -7.4 & 0.8 \\

\hline\hline
\end{tabular}
\tablefoot{We provide for each subgroup the number of stars used to compute the distance (after the RUWE filtering) and spatial velocity, Bayesian distance, mean, standard error of the mean (SEM), median, and standard deviation (SD) of the $UVW$ velocity components.}
}
\end{table*}

\subsection{Relative ages of the subgroups}\label{section4.5}

In this section we compare the relative age of the subgroups based on the location of the stars in the HR-diagram compared to evolutionary models and the fraction of disc-bearing stars in each population. First, we use the Virtual Observatory SED Analyzer \citep[VOSA,][]{VOSA} to fit the SED, derive the effective temperatures and bolometric luminosities of the stars. We use the BT-Settl \citep{Allard2012} grid of models to fit the SED and left the extinction $A_{V}$ as a free parameter (in the range of 0 to 10~mag). We use the optical photometry from Gaia-DR2 and infrared photometry from the 2MASS \citep{2MASS}, AllWISE \citep{Wright2010} and \textit{Spitzer} c2d \citep{Merin2008} surveys. We provide ourselves the photometric data of the stars to the VOSA service when possible to avoid erroneous cross-matches when querying these catalogues with the system interface. We cross-matched our sample of stars with the 2MASS and AllWISE catalogues using the source identifiers given in the auxiliary tables \texttt{TMASS\_BEST\_NEIGHBOUR} and \texttt{ALLWISE\_BEST\_NEIGHBOUR} available in the \textit{Gaia} archive. We derived effective temperatures and bolometric luminosities for 110~stars in the sample with available photometric data. The two stars of Lupus~2 in our sample are among the three rejected sources with insufficient photometric data to fit the SED with VOSA. 

We compare the effective temperatures obtained from the SED fits with those derived from the spectral type of the stars compiled from the literature \citep{Hughes1994,Krautter1997,Comeron2009,Comeron2013,Alcala2017}. We find spectral classification for only 69 stars (i.e. 61\% of the sample) and convert them to effective temperatures using the tables provided by \citet{Pecaut2013}. We observe a good agreement between the two effective temperature estimates with a mean deviation of $-32\pm31$~K (in the sense, result from spectral type `minus' result from SED fit) and rms of 260~K. We have therefore decided to use the effective temperature estimates derived from the SED fits in the following analysis to work with a homogeneous data set and use this information for a more significant number of stars in the sample.

\begin{figure*}[!h]
\begin{center}
\includegraphics[width=0.49\textwidth]{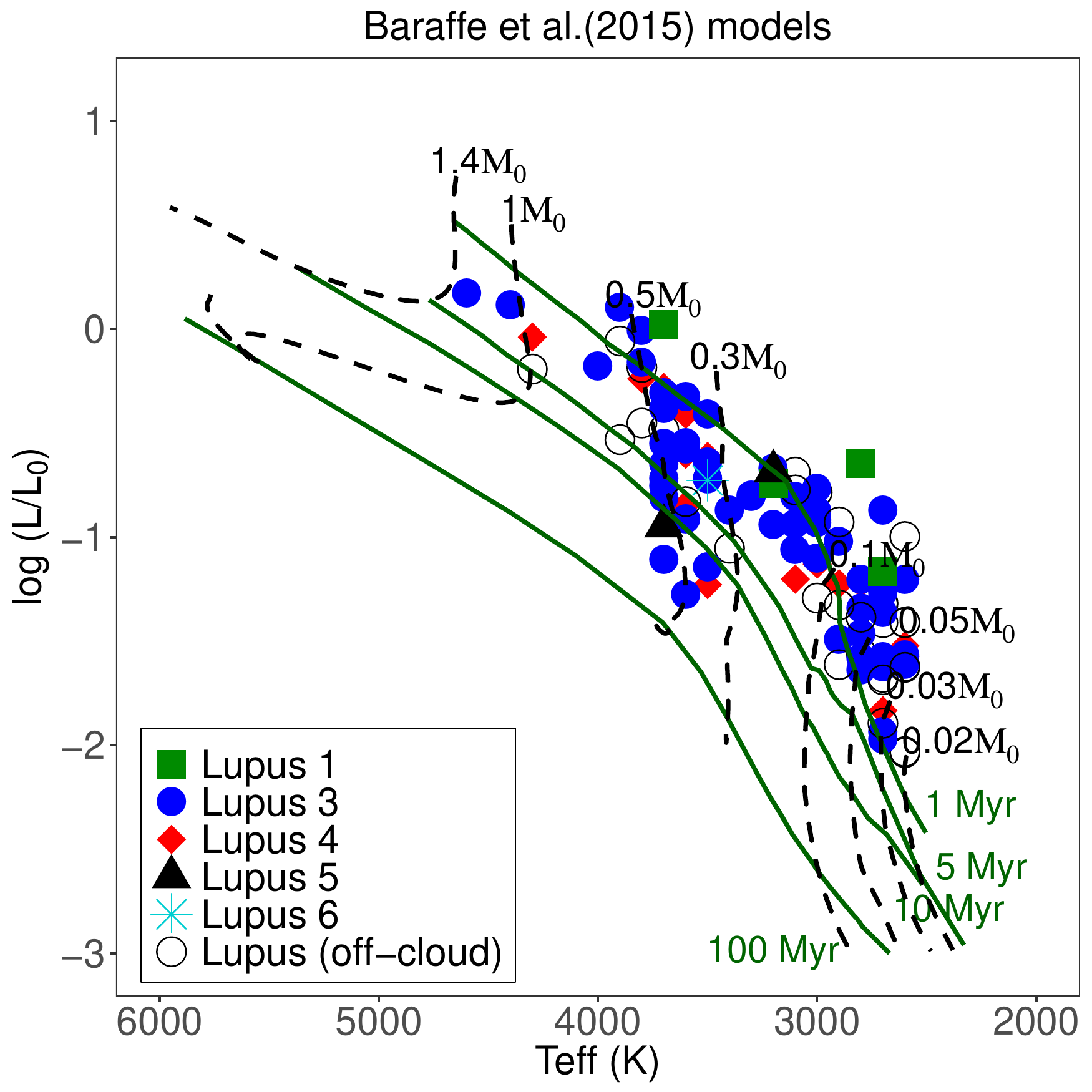}
\includegraphics[width=0.49\textwidth]{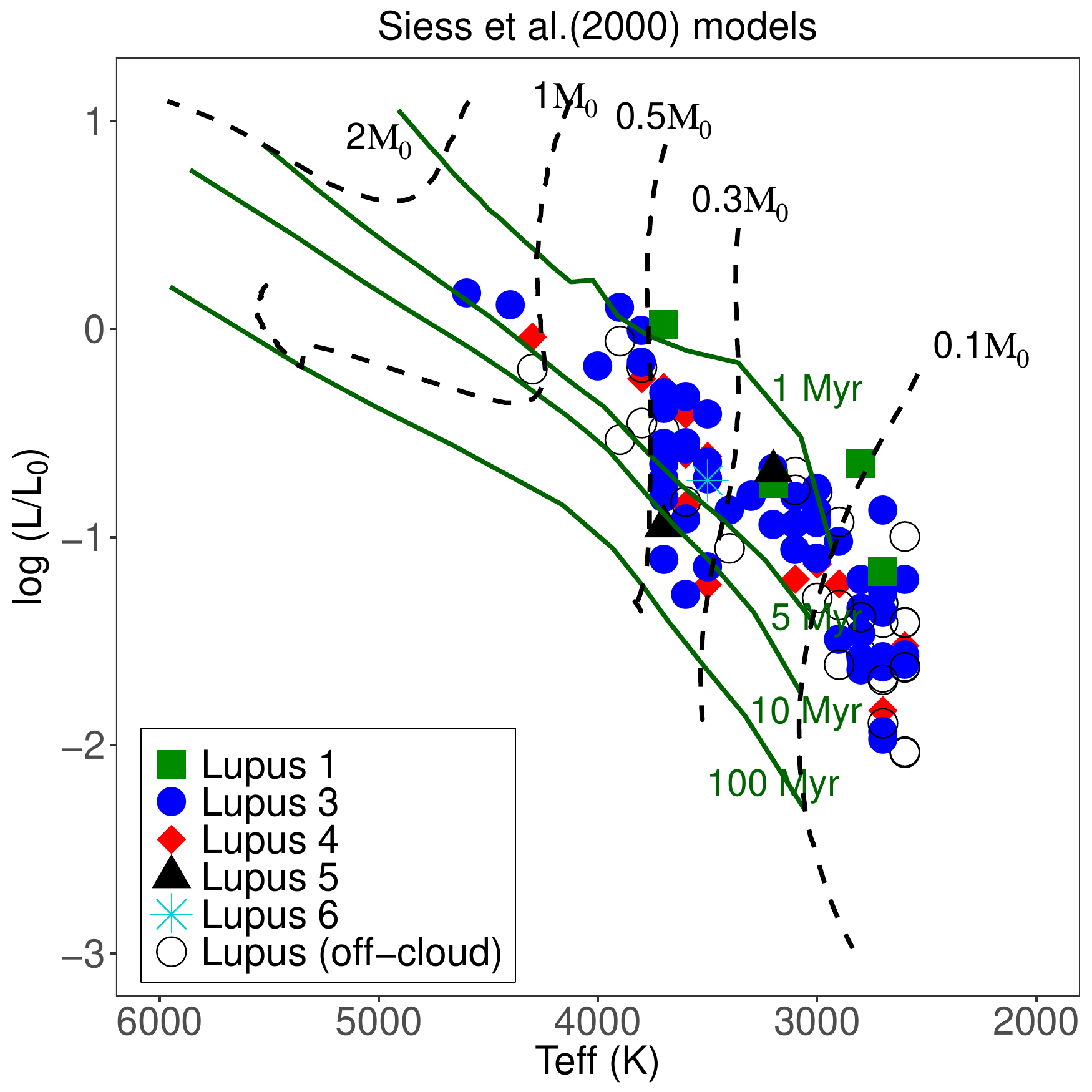}
\caption{HR diagram of the Lupus star-forming region with the \citet{BHAC15} and \citet{Siess2000} evolutionary models. The solid and dashed lines represent isochrones and tracks, respectively, with the corresponding ages (in Myr) and masses (in M$_{\odot}$) indicated in each panel. The most massive star in our sample (HR~5999) is not shown to improve the visibility of the low-mass range that includes the other stars in our sample. The error bars are smaller than the size of the symbols.
\label{fig_HRD} 
}
\end{center}
\end{figure*}

The resulting HR-diagram is shown in Figure~\ref{fig_HRD}. We compare the location of the stars in the HR-diagram with the \citet{BHAC15} and \citet{Siess2000} models that combined together cover the entire mass range of our sample which ranges from about 0.03 to 2.4M$_{\odot}$. Gaia~DR2~5997082081177906048 (HR~5999) is a binary early-type star and the most massive member in our sample \citep[$m=2.432\pm0.122$~M$_{\odot}$, see][]{Vioque2018}. The remaining sources in the sample are less massive than 1.4~M$_{\odot}$ and fall into the mass domain covered by the \citet{BHAC15} models. Our results suggest that the Lupus stars in the sample are mostly younger than 5~Myr. Many sources also lie above the 1~Myr isochrone and some of them are likely to be binaries or high-order multiple systems that will demand further investigation in future studies. Although age determination at these early stages of stellar evolution is rather uncertain, we detect no evident age gradient or segregation among the Lupus subgroups from the HR-diagram analysis. Table~\ref{tab_teff_age} provides the age estimates inferred from the \citet{BHAC15} and \citet{Siess2000} models together with the results obtained from the SED fit with VOSA.

Alternatively, we compare the age of the subgroups based on the fraction of disc-bearing stars in each population. We compute the spectral index $\alpha$ \citep{Lada1987} of each source in the sample with available photometric data in the infrared between 2 and 20~$\mu m$. This restricts the sample to 104~stars with AllWISE and Spitzer data. Then, we classify the stars into Class~I (embedded source, $\alpha>0$), Class~II (disc-bearing star, $-2<\alpha<0$) or Class~III (optically thin or no disc, $\alpha<-2$). We compare our results derived from the spectral index with the classification scheme proposed by \citet{Koenig2014} based on infrared colours and we obtain the same SED subclass for all stars in the sample. 

\begin{figure}
\begin{center}
\includegraphics[width=0.49\textwidth]{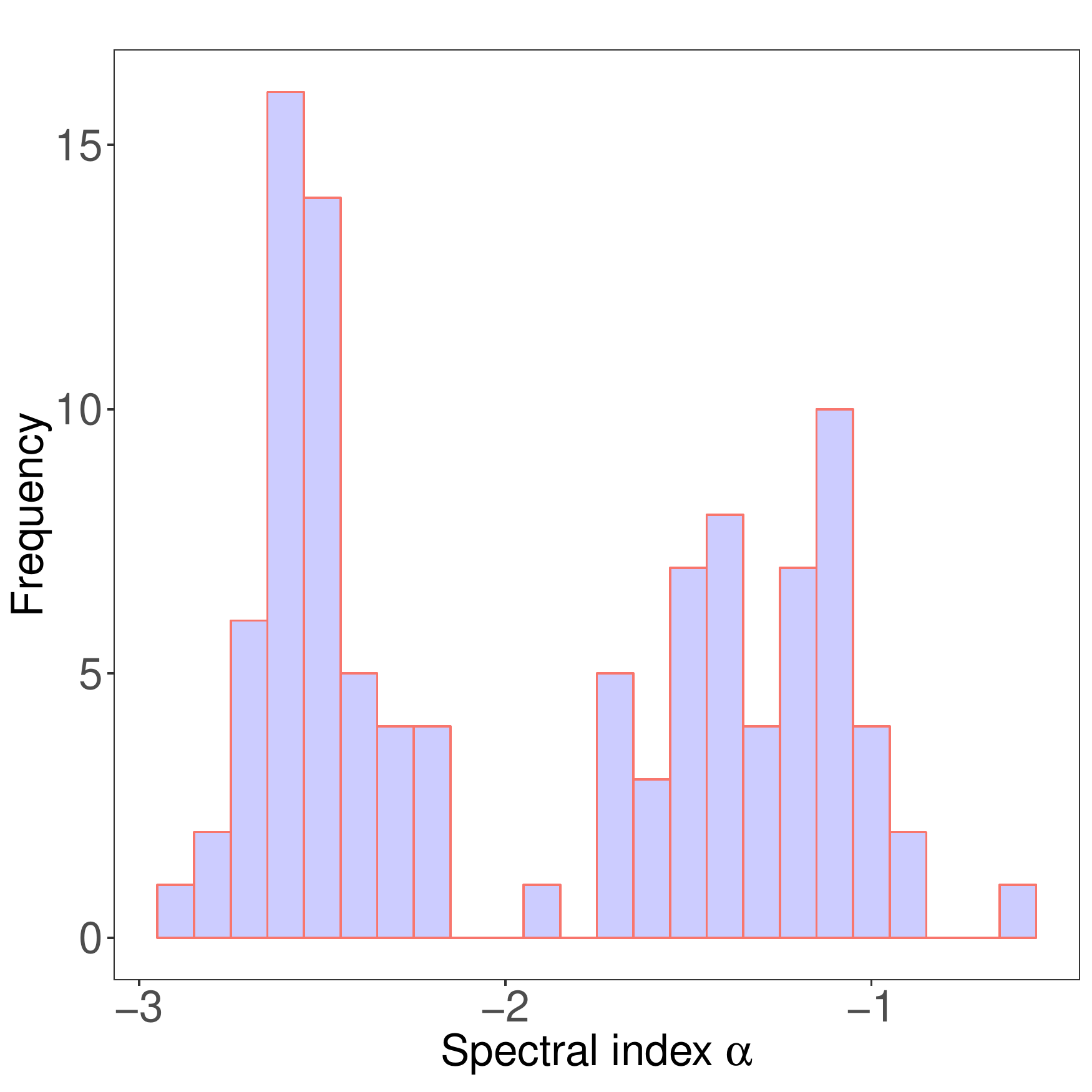}
\caption{Histogram of the spectral index $\alpha$ computed for 104 stars in the sample from infrared photometry.
\label{fig_alpha} 
}
\end{center}
\end{figure}

\begin{figure*}
\begin{center}
\includegraphics[width=0.49\textwidth]{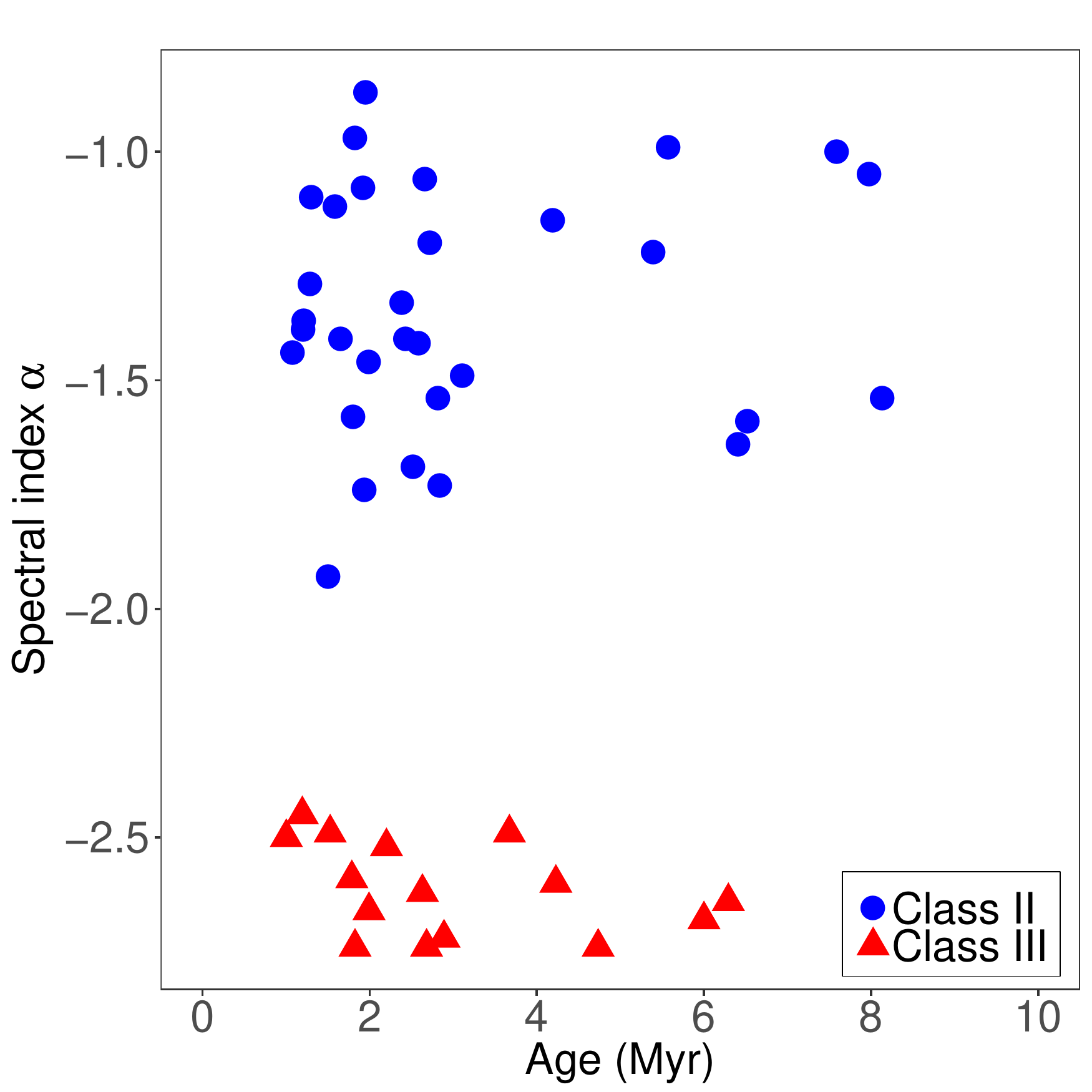}
\includegraphics[width=0.49\textwidth]{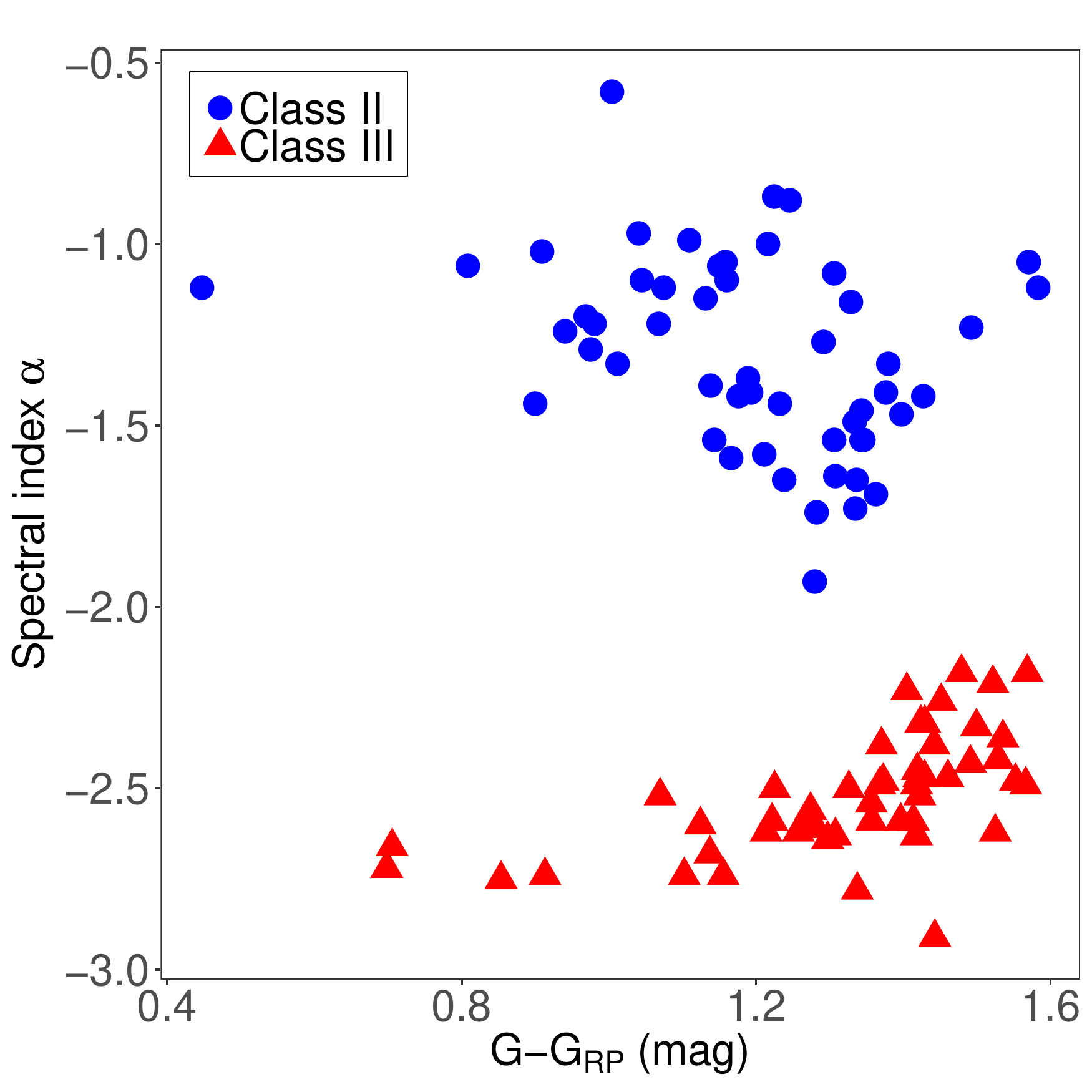}
\caption{Dependency of the spectral index on the age inferred from the \citet{BHAC15} models \textit{(left panel)} and colour of the stars using \textit{Gaia} photometry \textit{(right panel)}.
\label{fig_alpha_age} 
}
\end{center}
\end{figure*}

Figure~\ref{fig_alpha} shows that the distribution of the spectral index $\alpha$ (see Table~\ref{tab_members}) is bimodal suggesting the existence of two populations (disc-bearing and discless stars) with approximately the same number of sources. We verified that the bimodality of the distribution of spectral indices is not an artefact caused by the different number of points and wavelength range of the photometric data available for each star to compute the spectral indices. A similar result was also observed for the Chamaeleon~I star-forming region \citep[see Figure~11 of ][]{Luhman2008}. Previous studies suggested that the early disappearence of circumstellar discs could be related to environmental effects imposed by the presence of massive stars that produce strong UV radiation, stellar winds, and supernova explosions \citep[see e.g.][]{Walter1994,Martin1998}. However, we see no dependency of the spectral index on the position of the stars in our sample and any nearby OB star surrounding the Lupus clouds to support this scenario. As shown in Figure~\ref{fig_HRD} we note the existence of only a few stars older than 10~Myr in our sample which could be potential contaminants (as expected based on the performance of our classifier, see Table~\ref{tab_comp_pin}). Thus, the hypothesis of contamination by older field stars does not explain the bimodality of spectral indices in our sample given that the two populations have the same number of stars, similar ages and are younger than the potential contaminants from the Sco-Cen association. Alternatively, we investigated the dependency of the spectral indices on the age and colour of the stars. Figure~\ref{fig_alpha_age} shows that the age distribution of the Class~II and Class~III stars in Lupus overlap. The median age of Class~III stars inferred from the \citet{BHAC15} and \citet{Siess2000} stellar models is about 3~Myr which yields a rough estimate of the typical disc lifetime in the Lupus association. However, it should also be noted from Figure~\ref{fig_alpha_age} that we observe an excess of Class~III stars at cooler temperatures (red colours) suggesting that the survival time of circumstellar discs may also depend on other stellar parameters. For example, \citet{Galli2015} used an empirical disc evolution model to determine the lifetime of circumstellar discs in Lupus in terms of the mass of the star. According to their model the average lifetime of a circumstellar disc around a star with 0.1~M$_{\odot}$ is of the order of 1~Myr which could explain the early disapperance of circumstellar discs for some stars in our sample. 

\begin{figure*}[!h]
\begin{center}
\includegraphics[width=0.33\textwidth]{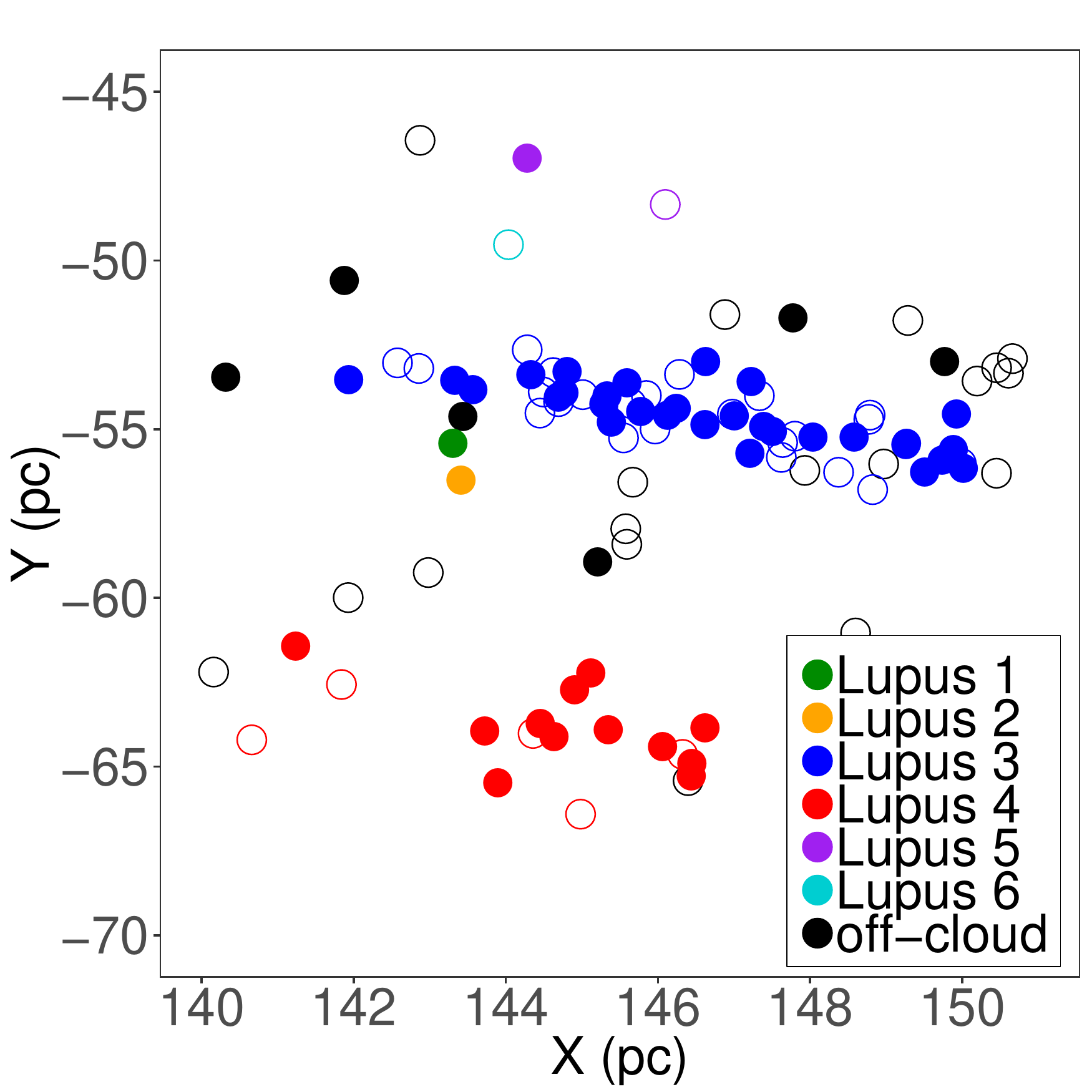}
\includegraphics[width=0.33\textwidth]{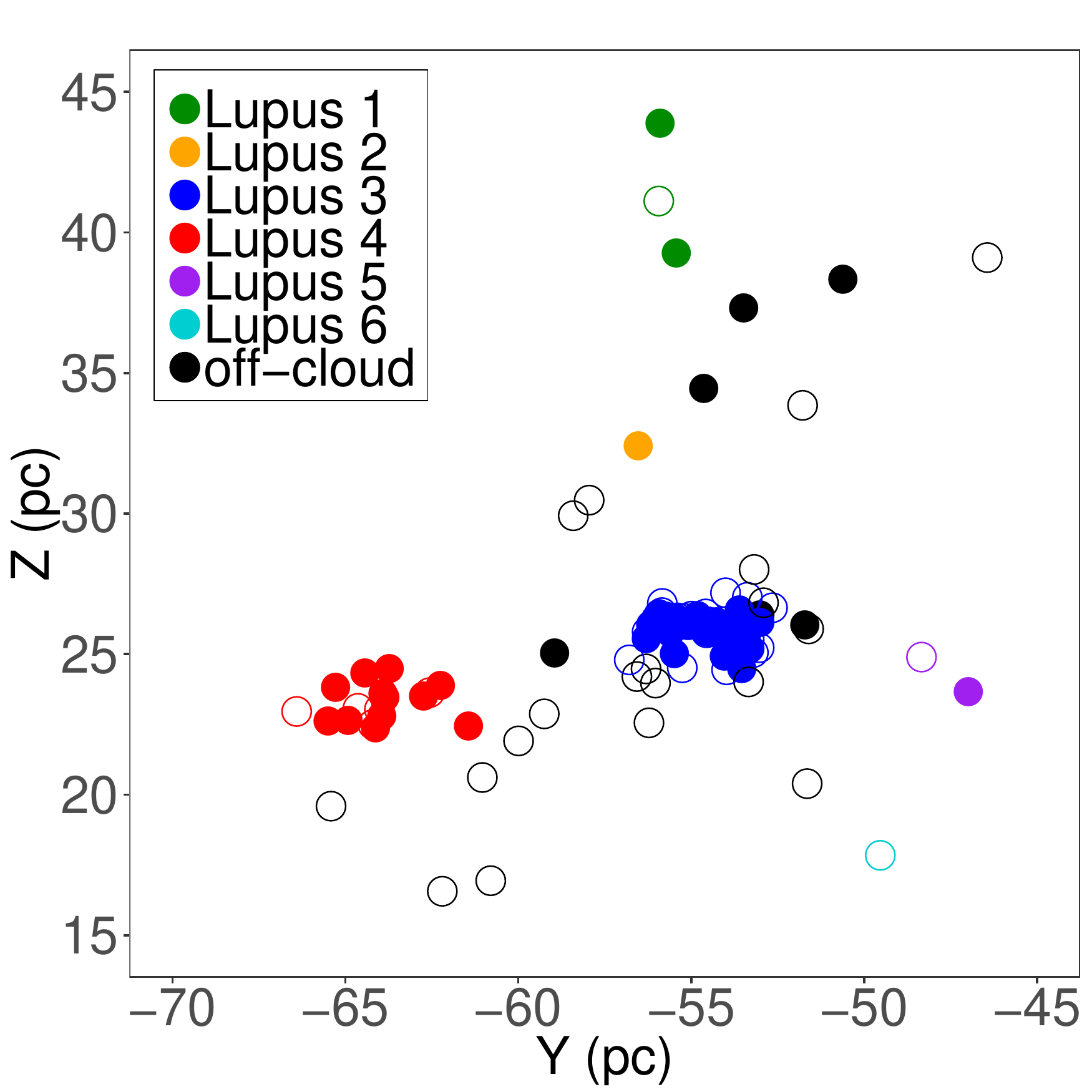}
\includegraphics[width=0.33\textwidth]{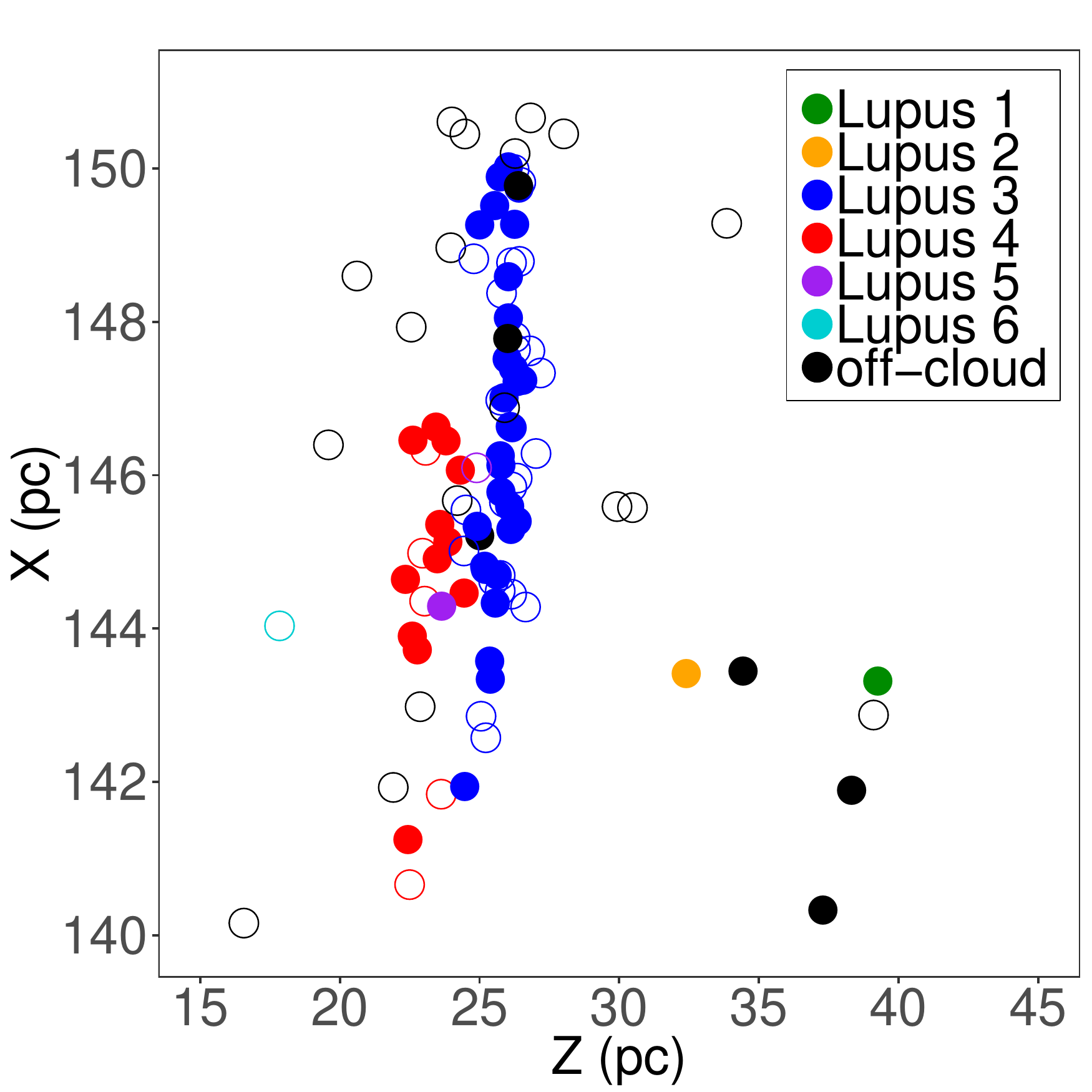}
\caption{Spatial distribution of the Lupus stars in our sample of cluster members. Colours indicate the different subgroups of the Lupus complex. Filled and open symbols denote Class~II and Class~III stars, respectively.
\label{fig_XYZ} 
}
\end{center}
\end{figure*}

In Table~\ref{tab_YSO} we compare the fraction of disc-bearing stars in the Lupus subgroups with their isochronal ages inferred from the \citet{BHAC15} and \citet{Siess2000} models. The following discussion about isochronal ages is obviously restricted to the stars covered by the evolutionary models used in this study (see Figure~\ref{fig_HRD}). This reduces the sample of stars with age estimates to only a few members in some subgroups, but the current analysis is still useful to provide a relative dating among them. For example, we note that Lupus~3 has similar fractions of disc-bearing and discless stars. As shown in Figure~\ref{fig_XYZ}, the populations of YSOs in Lupus~3 overlap in the 3D space of positions with no evident segregation between the two subclasses making it the most extended subgroup of the complex with about 8~pc in the $X$ direction (see also Figure~\ref{fig_3d}). It is interesting to note that the median ages of the Lupus subgroups inferred from isochrones range from about 1 to 3~Myr. The only exception shown in Table~\ref{tab_YSO} is Lupus~5, but our sample has only two stars. One of them, namely Gaia DR2 6021420630046381440, appears to be older than 10~Myr. It was first identified as a Lupus~5 member by \citet{Manara2018}, but the authors also raised the possibility that the targets selected in their study could be dispersed members of the Lupus~3 cloud. This would explain the discrepant age estimate that we observe for this star compared to the other source in our Lupus~5 sample, namely Gaia DR2 6021414479652887296, that has an age of about 1-2~Myr (see Figure~\ref{fig_HRD}) and is more consistent with the age reported for the other subgroups. We can therefore conclude that the Lupus subgroups are coeval. Interestingly, we also note that the off-cloud population of Lupus has similar ages compared to the on-cloud stars. The common age reinforces our previous conclusion in Sect.~\ref{sect4.4} based on spatial velocities that the off-cloud stars belong to the same association of the stars projected towards the Lupus clouds.

\begin{table}[!h]
\centering
\scriptsize{
\caption{Relative fraction of SED subclasses and age estimates for the Lupus subgroups.
\label{tab_YSO}}
\begin{tabular}{lccrrcccc}
\hline\hline
&&&&\multicolumn{2}{c}{BHAC15}&\multicolumn{2}{c}{SDF00}\\
&&&&\multicolumn{2}{c}{models}&\multicolumn{2}{c}{models}\\
\hline
Sample&$N_{\star}$&Class~II&Class~III&$N_{\star}$&Age&$N_{\star}$&Age \\
&&&&&(Myr)&&(Myr)\\
\hline\hline
Lupus~1&3&2 (67\%)&1 (33\%)&1&1.2&2&1.8\\
Lupus~2&1&1 (100\%)&0 (0\%)&0&\nodata&0&\nodata\\
Lupus~3&54&30 (56\%)&24 (44\%)&27&2.5&38&3.0\\
Lupus~4&17&12 (71\%)&5 (29\%)&13&2.4&13&3.7\\
Lupus~5&2&1 (50\%)&1 (50\%)&2&7.8&2&7.3\\
Lupus~6&1&0 (0\%)&1 (100\%)&1&2.7&1&3.7\\
Lupus~(off-cloud)&26&6 (23\%)&20 (77\%)&10&3.2&12&3.2\\
\hline
Lupus~(full sample)&104&52 (50\%)&52 (50\%)&54&2.6&68&3.1\\
\hline\hline
\end{tabular}
\tablefoot{We provide the number of stars and relative fraction of the SED subclasses in the parenthesis, number of stars with age estimate inferred from the \citet[][BHAC15]{BHAC15} and \citet[][SDF00]{Siess2000} models, and median age computed for each sample.}}
\end{table}

\subsection{Initial mass function}\label{section4.6}

In the following we use our new sample of members to discuss the initial mass function (IMF) of the Lupus association. As illustrated in Figure~\ref{fig_HRD} many stars in our sample fall into an age and mass domain that are not covered by the pre-main sequence stars evolutionary models published in the literature. We have therefore decided to use the distribution of spectral types in the sample as a proxy for the IMF. This procedure also allows us to compare our results in Lupus with other star-forming regions investigated in previous studies that adopted a similar approach.

We convert the effective temperatures derived from the SED fits (see Table~\ref{tab_teff_age}) to spectral types using the tables given by \citet{Pecaut2013}. We adopt the conversion from effective temperature to spectral type for 5-30~Myr old stars given in Table~6 of that study for most stars in our sample (with temperatures ranging from 2880 to 7280~K) and use the tabulated values for dwarf stars \citep[Table~5 of][]{Pecaut2013} for the remaining stars as an approximation. We estimate the resulting spectral types to be accurate by about 2~subclasses based on the rms of 260~K that we find when comparing the effective temperatures derived from the SED fit with the spectral classification of the stars with available information in our sample (see Sect.~\ref{section4.5}).

We estimate the completeness limits of our new sample of Lupus stars based on the completeness limits of the Gaia-DR2 catalogue which is the only source of data used for the membership analysis. The Gaia-DR2 catalogue is complete between $G=12$ and $G=17$~mag \citep[see e.g.][]{GaiaDR2}. The extinction is variable in the region covered by our survey (see Figure~\ref{fig_location_Lupus_final}) and the measurements given in the Gaia-DR2 catalogue for individual stars in our sample range from $A_{G}=0.8$ to $A_{G}=2.5$~mag with a median value of 1.5~mag. The Gaia-DR2 completeness limit of $G=17$~mag corrected for the maximum observed extinction of the stars in our sample at the distance of the Lupus clouds (see Table~\ref{tab_distance_velocity}) yields the absolute magnitude of $G_{abs}=8.5$~mag. This translates into 0.2~M$_{\odot}$ using the 3~Myr isochrone from the BT-Settl models \citep{Allard2012} and we therefore estimate our sample to be complete to this mass limit.

\begin{figure}[!h]
\begin{center}
\includegraphics[width=0.49\textwidth]{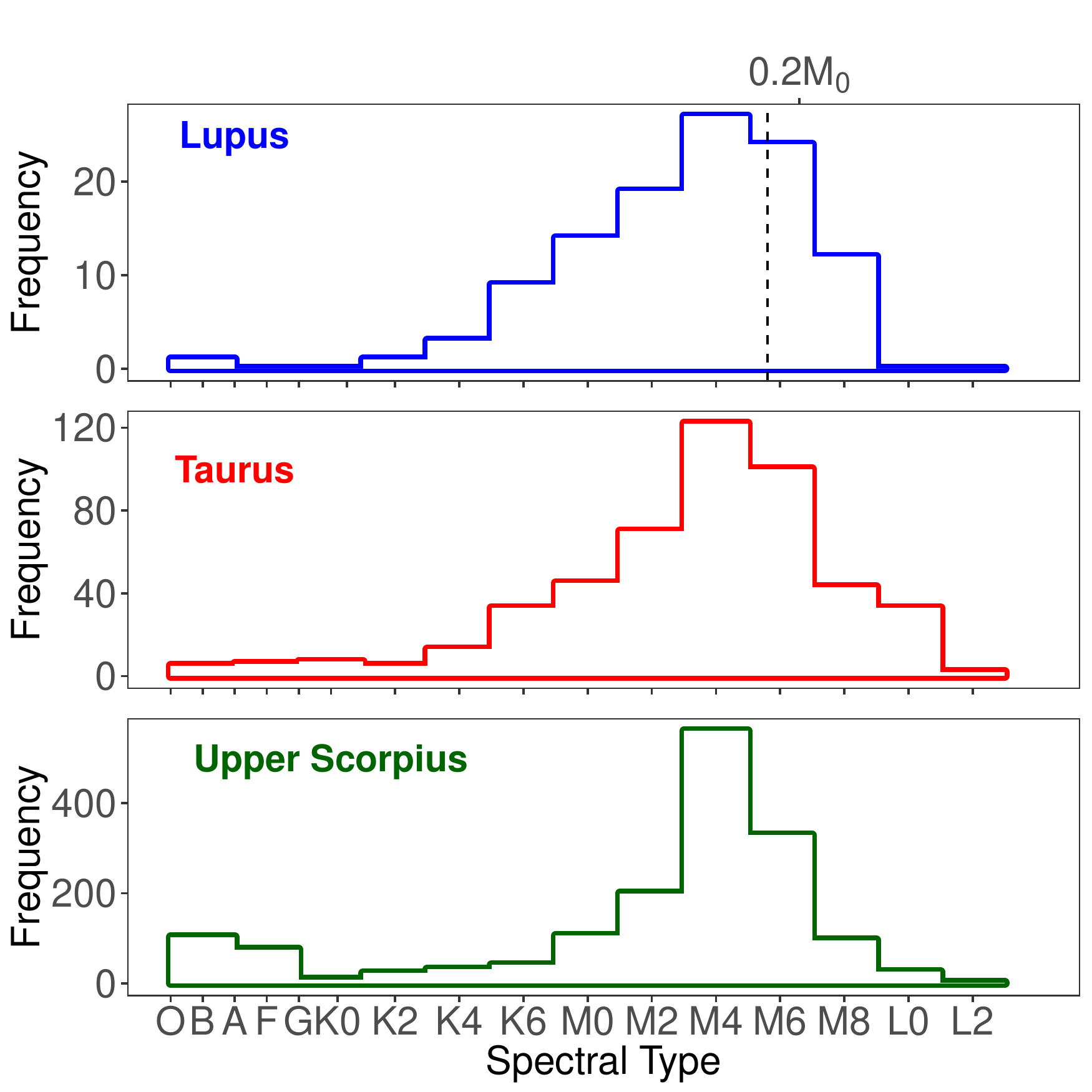}
\caption{Distribution of spectral types of the stars in Lupus (this paper), Taurus \citep{Esplin2019} and US \citep{Luhman2018} as a proxy for the IMF. The dashed line in the upper panel indicates the completeness limit (with the corresponding mass) of the Lupus sample investigated in this study.
\label{fig_IMF} 
}
\end{center}
\end{figure}

Figure~\ref{fig_IMF} shows the distribution of spectral types in the Lupus association compared to the Taurus and US star-forming regions using the spectral classifications derived by \citet{Esplin2019} and \citet{Luhman2018}, respectively. It is apparent that the sample of US exhibits a surplus of early-type stars which appear in small number in Taurus and are restricted to only one member (namely, HR~5999) in Lupus. Otherwise, we note that the distribution of spectral types in Lupus resembles the distribution of Taurus and US. We therefore argue that the IMF shows little variation for late-type stars among the three regions despite the different number of stars in the various samples. The samples of stars in Taurus and US show an important number of members with substellar masses which have been identified using ancillary data from optical and infrared surveys (in addition to the Gaia-DR2 catalogue). When we restrict the Taurus and US samples to the members that have been observed by the \textit{Gaia} satellite we find a similar shape for the distribution of spectral types at the faint end of the IMF that we obtain in this study for the Lupus association. 

Our study based on Gaia-DR2 data addressed many of the past uncertainties regarding membership, distance and the kinematic properties of the Lupus association returning the most complete census of stars down to about 0.2~M$_{\odot}$. This step was made necessary to further investigate the population of Lupus stars. Future studies using ancillary data to complement the \textit{Gaia} catalogue will search for the least-massive members and significantly increase the number of substellar objects. This will extend the IMF of the Lupus association down to the planetary-mass regime as done in other star-forming regions.  

\section{Conclusions}\label{section5}

In this study, we have revised the census of stars and properties of the various subgroups in the young stellar association located in the Lupus star-forming region based on Gaia-DR2 data. We applied a probabilistic method to infer membership probabilities of more than 10$^{7}$ sources over a field of 160~deg$^{2}$ encompassing the main star-forming clouds of the Lupus complex. We identified 137 stars (spread over the molecular clouds Lupus~1 to 6) that are probable members of the Lupus association. Our analysis confirms 90 stars from the literature that were previously associated with the Lupus region and adds 47 new members to list (an increase of more than 50\% with respect to the number of known members). We confirm that about 80\% of the historically known members of the region with available astrometry in the Gaia-DR2 catalogue are more likely to be background sources unrelated to Lupus, field stars or members of the adjacent Sco-Cen association in light of our new membership analysis. 

Our results on the 6D structure of the Lupus region show that the different subgroups of the complex (defined by the sample of stars projected towards the various molecular clouds) are located at the same distance (of about 160~pc) and move with the same spatial velocity. This confirms that the Lupus subgroups are comoving and belong to the same association. The HR-diagram shows that most stars in our sample are younger than 5~Myr and they cover the mass range from about 0.03 to 2.4$M_{\odot}$. Our results show that there is a superposition of ages between disc-bearing and discless stars in the Lupus association. The median age of the Lupus subgroups ranges from about 1 to 3~Myr and we therefore conclude that they are coeval. Both age indicators (fraction of disc-bearing stars and isochronal ages) reveal that the Lupus association is younger than the population of YSOs in the Corona-Australis star-forming region recently investigated by our team with the same methodology \citep{Galli2020}. Our new sample of Lupus stars is complete down to 0.2~M$_{\odot}$ and the IMF shows little variation compared to other star-forming regions.

The main limitation to investigate the dynamics of the Lupus complex (e.g. relative motion of the subgroups, expansion and rotation effects) is currently hampered by the precision on the radial velocities available in the literature (about 2~km/s). We therefore encourage astronomers to perform high-resolution spectroscopy of Lupus stars to allow for more precise studies on the kinematic properties of this region in conjunction with the future data releases of the \textit{Gaia} space mission.

\begin{acknowledgements}
We would like to thank the referee for carefully reading our manuscript and providing constructive comments. This research has received funding from the European Research Council (ERC) under the European Union’s Horizon 2020 research  and innovation programme (grant agreement No 682903, P.I. H. Bouy), and from the French State in the framework of the  ``Investments for the future'' Program, IdEx Bordeaux, reference ANR-10-IDEX-03-02. This research has made use of the SIMBAD database, operated at CDS, Strasbourg, France. This work has made use of data from the European Space Agency (ESA) mission {\it Gaia} (\url{https://www.cosmos.esa.int/gaia}), processed by the {\it Gaia} Data Processing and Analysis Consortium (DPAC, \url{https://www.cosmos.esa.int/web/gaia/dpac/consortium}). Funding for the DPAC has been provided by national institutions, in particular the institutions participating in the {\it Gaia} Multilateral Agreement. This publication makes use of data products from the Wide-field Infrared Survey Explorer, which is a joint project of the University of California, Los Angeles, and the Jet Propulsion Laboratory/California Institute of Technology, funded by the National Aeronautics and Space Administration. This publication makes use of VOSA, developed under the Spanish Virtual Observatory project supported by the Spanish MINECO through grant AyA2017-84089. VOSA has been partially updated by using funding from the European Union's Horizon 2020 Research and Innovation Programme, under Grant Agreement nº 776403 (EXOPLANETS-A) 
\end{acknowledgements}

\bibliographystyle{aa} 
\bibliography{references} 



\begin{appendix}
\section{Tables (online material)}\label{appendix_tables}

\begin{landscape}
\begin{table}
\caption{Properties of the 137 cluster members selected from our membership analysis in Lupus. (This table will be available in its entirety in machine-readable form.) }\label{tab_members}
\scriptsize{
\begin{tabular}{lccccccccccccccccc}
\hline\hline
Source Identifier&$\alpha$&$\delta$&$\mu_{\alpha}\cos\delta$&$\mu_{\delta}$&$\varpi$&RUWE&Prob.&$V_{r}$&Ref&$d$&$U$&$V$&$W$&Cloud&$\alpha$&SED\\
&(h:m:s) &($^{\circ}$ $^\prime$ $^\prime$$^\prime$)&(mas/yr)&(mas/yr)&(mas)&&&(km/s)&&(pc)&(km/s)&(km/s)&(km/s)&&&\\
\hline\hline

Gaia DR2 6013399894569703040 & 15 39 27.76 & -34 46 17.6 & $ -13.271 \pm 0.120 $& $ -22.242 \pm 0.069 $& $ 6.439 \pm 0.052 $& $ 1.02 $& $ 0.9307 $& $ -2.7 \pm 2.0 $& 1& $ 154.9 ^{+ 1.5 }_{ -1.3 } $& $ -7.3 ^{+ 1.9 }_{ -1.9 } $& $ -16.3 ^{+ 1.1 }_{ -1.1 } $& $ -7.1 ^{+ 1.0 }_{ -1.0 } $& Lupus 1 & $ -1.22 \pm 0.14 $& Class~II \\
Gaia DR2 6013073790593533824 & 15 44 00.94 & -35 31 06.1 & $ -11.492 \pm 0.264 $& $ -24.071 \pm 0.187 $& $ 6.454 \pm 0.140 $& $ 1.22 $& $ 0.9701 $& \nodata & \nodata & $ 156.0 ^{+ 2.7 }_{ -2.1 } $& \nodata & \nodata & \nodata & Lupus 1 & $ -2.47 \pm 0.11 $& Class~III \\
Gaia DR2 6014696841553696768 & 15 45 12.85 & -34 17 31.0 & $ -13.625 \pm 0.128 $& $ -21.605 \pm 0.081 $& $ 6.485 \pm 0.060 $& $ 1.83 $& $ 0.8220 $& \nodata & \nodata & $ 154.5 ^{+ 1.5 }_{ -1.1 } $& \nodata & \nodata & \nodata & Lupus 1 & \nodata & \nodata \\
Gaia DR2 6014769134444800896 & 15 46 42.97 & -34 30 11.9 & $ -12.575 \pm 0.395 $& $ -22.158 \pm 0.258 $& $ 6.088 \pm 0.214 $& $ 2.92 $& $ 0.8569 $& \nodata & \nodata & $ 159.9 ^{+ 2.5 }_{ -3.2 } $& \nodata & \nodata & \nodata & Lupus 1 & \nodata & \nodata \\
Gaia DR2 6007849461103723136 & 15 47 11.58 & -41  01 18.5 & $ -13.475 \pm 0.167 $& $ -22.863 \pm 0.115 $& $ 6.405 \pm 0.082 $& $ 2.05 $& $ 0.8716 $& \nodata & \nodata & $ 155.9 ^{+ 2.1 }_{ -1.8 } $& \nodata & \nodata & \nodata & off-cloud & \nodata & \nodata \\
Gaia DR2 6011581856393988352 & 15 48 06.24 & -35 15 48.5 & $ -12.124 \pm 0.195 $& $ -22.325 \pm 0.129 $& $ 6.607 \pm 0.093 $& $ 1.31 $& $ 0.9461 $& \nodata & \nodata & $ 154.0 ^{+ 1.4 }_{ -1.0 } $& \nodata & \nodata & \nodata & Lupus 1 & \nodata & \nodata \\
Gaia DR2 6011500389453302272 & 15 49 30.72 & -35 49 51.8 & $ -12.767 \pm 0.110 $& $ -23.370 \pm 0.077 $& $ 6.268 \pm 0.051 $& $ 1.07 $& $ 0.9620 $& $ 1.4 \pm 1.0 $& 1& $ 158.6 ^{+ 1.8 }_{ -1.7 } $& $ -3.8 ^{+ 1.0 }_{ -1.0 } $& $ -18.6 ^{+ 0.8 }_{ -0.8 } $& $ -6.7 ^{+ 0.7 }_{ -0.7 } $& Lupus 1 & $ -1.37 \pm 0.27 $& Class~II \\
Gaia DR2 5995437246144906496 & 15 51 31.67 & -43  02 04.8 & $ -8.830 \pm 0.977 $& $ -22.605 \pm 0.671 $& $ 5.577 \pm 0.487 $& $ 10.94 $& $ 0.8366 $& \nodata & \nodata & $ 159.7 ^{+ 2.8 }_{ -4.0 } $& \nodata & \nodata & \nodata & off-cloud & \nodata & \nodata \\
Gaia DR2 6011392186336443392 & 15 51 46.94 & -35 56 44.5 & $ -12.418 \pm 0.088 $& $ -24.158 \pm 0.058 $& $ 6.459 \pm 0.052 $& $ 1.15 $& $ 0.9602 $& $ -2.6 \pm 0.1 $& 2& $ 154.7 ^{+ 1.5 }_{ -1.2 } $& $ -7.2 ^{+ 0.2 }_{ -0.2 } $& $ -17.0 ^{+ 0.4 }_{ -0.5 } $& $ -7.8 ^{+ 0.5 }_{ -0.5 } $& off-cloud & $ -1.24 \pm 0.28 $& Class~II \\
Gaia DR2 6011827867821601792 & 15 55 10.26 & -34 55 05.0 & $ -11.096 \pm 0.535 $& $ -23.941 \pm 0.306 $& $ 6.778 \pm 0.257 $& $ 0.96 $& $ 0.9191 $& \nodata & \nodata & $ 155.4 ^{+ 3.2 }_{ -1.9 } $& \nodata & \nodata & \nodata & off-cloud & $ -1.12 \pm 0.13 $& Class~II \\
Gaia DR2 6010133559067032832 & 15 55 50.27 & -38  01 34.1 & $ -11.659 \pm 0.129 $& $ -22.951 \pm 0.078 $& $ 6.255 \pm 0.061 $& $ 1.18 $& $ 0.9961 $& $ -0.1 \pm 2.9 $& 1& $ 158.9 ^{+ 1.8 }_{ -1.9 } $& $ -5.3 ^{+ 2.8 }_{ -2.8 } $& $ -17.3 ^{+ 1.5 }_{ -1.4 } $& $ -7.1 ^{+ 1.0 }_{ -1.0 } $& Lupus 2 & \nodata & \nodata \\
Gaia DR2 6010133559067032704 & 15 55 50.32 & -38  01 32.2 & $ -11.787 \pm 0.173 $& $ -23.359 \pm 0.103 $& $ 6.249 \pm 0.080 $& $ 2.70 $& $ 0.9034 $& \nodata & \nodata & $ 159.0 ^{+ 2.2 }_{ -2.2 } $& \nodata & \nodata & \nodata & Lupus 2 & \nodata & \nodata \\
Gaia DR2 6010483616079976448 & 15 56 02.08 & -36 55 28.6 & $ -11.660 \pm 0.073 $& $ -22.503 \pm 0.047 $& $ 6.329 \pm 0.037 $& $ 1.03 $& $ 0.9918 $& $ 2.6 \pm 1.2 $& 1& $ 157.3 ^{+ 1.4 }_{ -1.4 } $& $ -2.4 ^{+ 1.2 }_{ -1.2 } $& $ -17.9 ^{+ 0.8 }_{ -0.8 } $& $ -6.1 ^{+ 0.6 }_{ -0.7 } $& off-cloud & $ -1.20 \pm 0.14 $& Class~II \\
Gaia DR2 6010135758090335232 & 15 56 09.19 & -37 56 06.5 & $ -12.091 \pm 0.120 $& $ -23.718 \pm 0.074 $& $ 6.311 \pm 0.054 $& $ 1.27 $& $ 0.9957 $& $ -1.0 \pm 1.0 $& 3& $ 157.5 ^{+ 1.8 }_{ -1.8 } $& $ -6.2 ^{+ 1.1 }_{ -1.1 } $& $ -17.4 ^{+ 0.8 }_{ -0.8 } $& $ -7.4 ^{+ 0.7 }_{ -0.7 } $& Lupus 2 & $ -1.02 \pm 0.13 $& Class~II \\
Gaia DR2 5994793310284482432 & 15 56 37.72 & -42 42 45.0 & $ -10.443 \pm 0.369 $& $ -22.872 \pm 0.249 $& $ 5.986 \pm 0.175 $& $ 1.01 $& $ 0.8959 $& \nodata & \nodata & $ 161.1 ^{+ 1.9 }_{ -2.7 } $& \nodata & \nodata & \nodata & Lupus 4 & $ -2.48 \pm 0.12 $& Class~III \\
Gaia DR2 5994795990344219904 & 15 56 38.11 & -42 35 57.6 & $ -12.117 \pm 0.304 $& $ -23.561 \pm 0.212 $& $ 6.420 \pm 0.145 $& $ 1.04 $& $ 0.9352 $& \nodata & \nodata & $ 156.2 ^{+ 3.2 }_{ -2.3 } $& \nodata & \nodata & \nodata & Lupus 4 & $ -2.26 \pm 0.05 $& Class~III \\
Gaia DR2 6010114558131195392 & 15 56 42.30 & -37 49 15.8 & $ -11.546 \pm 0.139 $& $ -23.234 \pm 0.090 $& $ 6.267 \pm 0.067 $& $ 1.80 $& $ 0.9966 $& \nodata & \nodata & $ 158.8 ^{+ 2.0 }_{ -2.0 } $& \nodata & \nodata & \nodata & Lupus 2 & \nodata & \nodata \\
Gaia DR2 5994793241564999040 & 15 56 44.20 & -42 42 24.9 & $ -10.973 \pm 0.197 $& $ -22.833 \pm 0.133 $& $ 6.140 \pm 0.097 $& $ 1.13 $& $ 0.9709 $& \nodata & \nodata & $ 161.0 ^{+ 1.7 }_{ -2.3 } $& \nodata & \nodata & \nodata & Lupus 4 & \nodata & \nodata \\
Gaia DR2 5994747367001754240 & 15 57 23.99 & -42 40 04.9 & $ -11.259 \pm 0.075 $& $ -23.205 \pm 0.052 $& $ 6.229 \pm 0.035 $& $ 0.98 $& $ 0.9962 $& \nodata & \nodata & $ 159.7 ^{+ 1.5 }_{ -1.4 } $& \nodata & \nodata & \nodata & Lupus 4 & $ -1.10 \pm 0.14 $& Class~II \\
Gaia DR2 5994831449595857408 & 15 57 30.32 & -42 10 32.8 & $ -11.671 \pm 0.109 $& $ -23.292 \pm 0.076 $& $ 6.204 \pm 0.052 $& $ 1.42 $& $ 0.9931 $& \nodata & \nodata & $ 160.4 ^{+ 1.6 }_{ -1.6 } $& \nodata & \nodata & \nodata & Lupus 4 & \nodata & \nodata \\

\hline
\hline

\end{tabular}
\tablefoot{For each star, we provide the Gaia-DR2 identifier, position, proper motion and parallax (not corrected for zero-point offset) from the Gaia-DR2 catalogue, RUWE, membership probability, radial velocity with reference, distance derived from Bayesian inference, UVW spatial velocity, molecular cloud, spectral index, and object class based on the SED. References for radial velocities: (1)~\citet{Frasca2017}, (2)~\citet{Guenther2007}, (3)~\citet{Wichmann1999}, (4)~\citet{Gontcharov2006}, and (5)~\citet{Galli2013}.}
}
\end{table}
\end{landscape}
\clearpage
----------------------------------------------------
\begin{table*}
\centering
\caption{Membership probability for all sources in the field derived independently using different probability threshold values for $p_{in}$. (This table will be available in its entirety in machine-readable form.)
\label{tab_prob}}
\begin{tabular}{cccccc}
\hline\hline
Source Identifier&probability&probability&probability&probability&probability\\
&($p_{in}=0.5$)&($p_{in}=0.6$)&($p_{in}=0.7$)&($p_{in}=0.8$)&($p_{in}=0.9$)\\
\hline\hline

Gaia DR2 6211958981441554048	&	1.77E-47	&	1.91E-44	&	1.08E-45	&	3.24E-46	&	1.92E-78	\\
Gaia DR2 6211958985738882560	&	2.55E-261	&	3.15E-272	&	2.88E-284	&	7.31E-298	&	7.31E-298	\\
Gaia DR2 6211958740923365760	&	5.31E-294	&	5.31E-294	&	5.31E-294	&	5.31E-294	&	5.31E-294	\\
Gaia DR2 6211958779577618944	&	8.57E-84	&	5.80E-82	&	2.97E-83	&	1.69E-83	&	9.93E-124	\\
Gaia DR2 6211958745217873024	&	3.78E-86	&	2.01E-85	&	4.34E-87	&	1.33E-87	&	1.37E-110	\\
Gaia DR2 6211958981441557120	&	3.00E-72	&	3.28E-70	&	1.70E-71	&	7.88E-72	&	6.99E-104	\\
Gaia DR2 6211958779580451584	&	6.46E-298	&	6.46E-298	&	6.46E-298	&	6.46E-298	&	6.46E-298	\\
Gaia DR2 6211958775283118720	&	5.24E-104	&	1.88E-109	&	4.49E-112	&	7.97E-116	&	7.50E-173	\\
Gaia DR2 6211958775283123712	&	3.53E-125	&	3.78E-125	&	6.47E-130	&	1.83E-132	&	8.38E-180	\\
Gaia DR2 6211959088818099072	&	3.89E-297	&	3.89E-297	&	3.89E-297	&	3.89E-297	&	3.89E-297	\\
Gaia DR2 6211958676498395136	&	2.59E-81	&	1.70E-80	&	1.24E-81	&	5.98E-82	&	1.20E-125	\\
Gaia DR2 6211958706563631488	&	8.05E-167	&	1.28E-170	&	3.72E-177	&	9.72E-180	&	1.90E-253	\\
Gaia DR2 6211958676498394240	&	3.98E-83	&	1.14E-81	&	3.08E-84	&	6.01E-85	&	2.38E-120	\\
Gaia DR2 6211958882659668352	&	4.90E-296	&	4.90E-296	&	4.90E-296	&	4.90E-296	&	4.90E-296	\\
Gaia DR2 6211957916289638784	&	1.56E-285	&	1.00E-293	&	9.89E-298	&	9.89E-298	&	9.89E-298	\\
Gaia DR2 6211958710858136192	&	8.24E-94	&	1.16E-92	&	1.35E-95	&	2.11E-96	&	1.55E-133	\\
Gaia DR2 6211957916289619072	&	1.24E-296	&	1.24E-296	&	1.24E-296	&	1.24E-296	&	1.24E-296	\\
Gaia DR2 6211958882659669120	&	4.23E-297	&	4.23E-297	&	4.23E-297	&	4.23E-297	&	4.23E-297	\\
Gaia DR2 6211958882656846208	&	7.73E-118	&	3.89E-119	&	2.10E-121	&	8.46E-123	&	5.50E-172	\\
Gaia DR2 6211958878362349184	&	7.35E-206	&	2.84E-212	&	1.55E-217	&	1.01E-221	&	3.09E-294	\\

\hline\hline
\end{tabular}
\end{table*}

\begin{table*}
\centering
\caption{Empirical isochrone of the Lupus association inferred from our membership analysis. (This table will be available in its entirety in machine-readable form.)
\label{tab_isochrone}}
\begin{tabular}{cc}
\hline\hline
$G_{RP}$&$G-G_{RP}$\\
(mag)&(mag)\\
\hline\hline
7.331&0.425\\
7.356&0.429\\
7.381&0.432\\
7.406&0.435\\
7.431&0.439\\
7.457&0.442\\
7.482&0.445\\
7.507&0.449\\
7.532&0.452\\
7.557&0.455\\
7.583&0.459\\
7.608&0.462\\
7.633&0.465\\
7.658&0.469\\
7.683&0.472\\
7.709&0.475\\
7.734&0.479\\
7.759&0.482\\
7.784&0.485\\
7.809&0.489\\
\hline\hline
\end{tabular}
\end{table*}

\begin{table*}
\centering
\caption{Stellar parameters for the sample of 110 members with available photometry for the SED analysis. (This table will be available in its entirety in machine-readable form.)
\label{tab_teff_age}}
\begin{tabular}{cccccc}
\hline\hline
Source Identifier&$T_{eff}$&$\log L$&$A_{V}$&$t_{BHAC15}$&$t_{SDF00}$\\
&(K)&(L in $L_{\odot}$)&(mag)&(Myr)&(Myr)\\
\hline\hline

Gaia DR2 6013399894569703040 & $ 3700 \pm 50 $& $ +0.023 ^{+ 0.014 }_{ -0.015 } $& 0.5 & & $ 0.9 ^{+ 0.1 }_{ -0.1 } $\\
Gaia DR2 6013073790593533824 & $ 2700 \pm 50 $& $ -1.163 ^{+ 0.012 }_{ -0.013 } $& 0.5 & & \\
Gaia DR2 6011581856393988352 & $ 2800 \pm 50 $& $ -0.646 ^{+ 0.010 }_{ -0.011 } $& 0.5 & & \\
Gaia DR2 6011500389453302272 & $ 3200 \pm 50 $& $ -0.743 ^{+ 0.012 }_{ -0.012 } $& 0.5 & $ 1.2 ^{+ 0.2 }_{ -0.2 } $& $ 2.6 ^{+ 0.1 }_{ -0.1 } $\\
Gaia DR2 6011392186336443392 & $ 3900 \pm 50 $& $ -0.058 ^{+ 0.012 }_{ -0.013 } $& 0.5 & & $ 1.5 ^{+ 0.3 }_{ -0.2 } $\\
Gaia DR2 6011827867821601792 & $ 2700 \pm 50 $& $ -1.895 ^{+ 0.013 }_{ -0.014 } $& 1.0 & & \\
Gaia DR2 6010483616079976448 & $ 3800 \pm 50 $& $ -0.452 ^{+ 0.013 }_{ -0.014 } $& 0.5 & $ 2.7 ^{+ 0.7 }_{ -0.6 } $& $ 3.5 ^{+ 0.8 }_{ -0.6 } $\\
Gaia DR2 5994793310284482432 & $ 2600 \pm 50 $& $ -1.574 ^{+ 0.012 }_{ -0.013 } $& 1.0 & & \\
Gaia DR2 5994795990344219904 & $ 2600 \pm 50 $& $ -1.522 ^{+ 0.013 }_{ -0.013 } $& 0.5 & & \\
Gaia DR2 5994793241564999040 & $ 3500 \pm 50 $& $ -1.228 ^{+ 0.017 }_{ -0.017 } $& 1.0 & $ 18.7 ^{+ 5.8 }_{ -4.7 } $& $ 14.3 ^{+ 4.0 }_{ -2.6 } $\\
Gaia DR2 5994747367001754240 & $ 3600 \pm 50 $& $ -0.411 ^{+ 0.012 }_{ -0.013 } $& 0.5 & $ 1.3 ^{+ 0.3 }_{ -0.2 } $& $ 1.9 ^{+ 0.3 }_{ -0.1 } $\\
Gaia DR2 5995233114932232064 & $ 3700 \pm 50 $& $ -0.281 ^{+ 0.012 }_{ -0.012 } $& 0.0 & $ 1.1 ^{+ 0.3 }_{ 0.0 } $& $ 1.8 ^{+ 0.1 }_{ -0.1 } $\\
Gaia DR2 5995206142536938112 & $ 2800 \pm 50 $& $ -1.223 ^{+ 0.012 }_{ -0.012 } $& 0.5 & & \\
Gaia DR2 5995219680274445184 & $ 3500 \pm 50 $& $ -0.612 ^{+ 0.013 }_{ -0.014 } $& 0.0 & $ 1.8 ^{+ 0.4 }_{ -0.3 } $& $ 2.8 ^{+ 0.3 }_{ -0.3 } $\\
Gaia DR2 5995219680274444672 & $ 3100 \pm 50 $& $ -1.202 ^{+ 0.013 }_{ -0.013 } $& 0.5 & $ 2.8 ^{+ 0.7 }_{ -0.5 } $& $ 4.6 ^{+ 0.5 }_{ -0.4 } $\\
Gaia DR2 5998031578183659776 & $ 2800 \pm 50 $& $ -1.385 ^{+ 0.013 }_{ -0.013 } $& 0.5 & & \\
Gaia DR2 5997975400009713920 & $ 2600 \pm 50 $& $ -1.626 ^{+ 0.013 }_{ -0.013 } $& 0.5 & & \\
Gaia DR2 5995154907858605312 & $ 3600 \pm 50 $& $ -0.602 ^{+ 0.015 }_{ -0.015 } $& 0.5 & $ 2.4 ^{+ 0.6 }_{ -0.5 } $& $ 3.1 ^{+ 0.6 }_{ -0.3 } $\\
Gaia DR2 5995157042469914752 & $ 2700 \pm 50 $& $ -1.833 ^{+ 0.013 }_{ -0.014 } $& 0.5 & & \\
Gaia DR2 5995168724780802944 & $ 3800 \pm 50 $& $ -0.239 ^{+ 0.014 }_{ -0.014 } $& 0.5 & $ 1.3 ^{+ 0.3 }_{ -0.3 } $& $ 1.9 ^{+ 0.3 }_{ -0.2 } $\\

\hline\hline
\end{tabular}
\tablefoot{We provide for each star its effective temperature and bolometric luminosity derived from the SED fit with VOSA, the extinction value for dereddening the SED, and the ages inferred from the \citet[][BHAC15]{BHAC15} and \citet[][SDF00]{Siess2000} models.
}
\end{table*}

\end{appendix}
\end{document}